**Title:**

# Fast Programming of In-Plane Hyperbolic Phonon Polariton Optics Through van der Waals Crystals using the Phase-Change Material $In_3SbTe_2$

Author(s) and Corresponding Author(s)*:

Lina Jäckering*,+,[1,2], Umberto Saldarelli+,[1,2], Aaron Moos[1,2], Lukas Conrads[1,2], Enrique Terán-García[3,4], Christian Lanza[3,4], Aitana Tarazaga Martín-Luengo[3,4], Gonzalo Álvarez-Pérez[5], Pablo Alonso-González[3,4], Matthias Wuttig[1,2], Thomas Taubner*,[1,2]

jaeckering@physik.rwth-aachen.de

+both authors contributed equally

**Affiliations**

[1]1st Institute of Physics (IA), RWTH Aachen University, 52074 Aachen, Germany

[2]Juelich-Aachen Research Alliance (JARA-FIT), 52425 Juelich, Germany

[3]Department of Physics, University of Oviedo, 33006 Oviedo, Spain

[4]Center of Research on Nanomaterials and Nanotechnology, CINN (CSIC-Universidad de Oviedo), 33940 El Entrego, Spain

[5]Istituto Italiano di Tecnologia, Center for Biomolecular Nanotechnologies, Via Barsanti 14, 73010 Arnesano, Italy.

**Abstract (197/200)**

The high directionality of hyperbolic phonon polaritons (HPhPs) has opened radically new ways to route and steer the flow of energy at the nanoscale. However, launching HPhPs requires fabricating efficient and precisely aligned polariton launching structures, which remains time-consuming and expensive with conventional nanofabrication approaches. Recently, using optical laser pulses, polariton launching structures have been programmed into the plasmonic phase-change material $In_3SbTe_2$. Here, we leverage this approach to reconfigure HPhPs by programming a variety of launching and confining nanostructures through α-$MoO_3$ flakes deposited onto $In_3SbTe_2$. Importantly, optical programming after flake deposition enables alignment of launching stripes to the [001]-axis of the flake, essential to control the directional polariton propagation. We showcase these capabilities in a variety of structures: i) an optically programmed disk, showing similar tuning ranges and confinement as focusing by gold disks; and ii) a cavity for in-plane HPhPs created by reconfiguring the single disk to a double disk structure, tailoring the confinement by simply reprogramming the disk distance. Our fabrication scheme offers fast turn-around times, flexible alignment and the opportunity to reconfigure the structures. Thus, it is a fast, efficient and versatile way to tailor propagation and confinement of highly directional polaritons on demand.

## Introduction

Polaritons – hybridized excitations of photons and a collective, dipolar oscillation in a material – in van der Waals (vdW) materials are prone to high confinement and therefore prime candidates for nanophotonic devices. VdW materials host a plethora of polaritons of different kinds[1], among those surface plasmon polaritons in graphene[2,3], out-of-plane hyperbolic phonon polaritons in hBN[4], and in-plane hyperbolic phonon polaritons in $\alpha$-MoO$_3$[5–7], in $\alpha$-V$_2$O$_5$[8] or in LiV$_2$O$_5$[9]. The in-plane hyperbolic phonon polaritons in $\alpha$-MoO$_3$ have been demonstrated to show high directionality[10–12], planar and negative refraction[13–16], hyperfocusing[14,17–21] and negative reflection[22]. These unusual, fundamental optical phenomena at the nanoscale[23] result from the high directionality of the polaritons in $\alpha$-MoO$_3$. This exceptional high directionality facilitates routing and steering the polaritons and thereby the flow of energy at the nanoscale which makes the highly directional polaritons promising for nanophotonic devices[24].

Steering polaritons and studying fundamental optical phenomena in anisotropic vdW materials requires fabrication of optical structures with precise angular alignment relative to the crystal axes. Conventionally, these optical structures are realized by adding metallic launching structures or by etching away parts of the polariton hosting material.[5,10–14,17–21] However, these conventional tailoring approaches rely on cumbersome and time-consuming fabrication techniques, including lithography and etching (c.f. Supplementary Note 1). Especially, the fabrication of launching structures to explore in-plane anisotropic materials such as $\alpha$-MoO$_3$ requires precise alignment of the vdW material flake with the fabricated structure as their alignment is crucial for controlling the in-plane anisotropic propagation. Further, structures realized through conventional fabrication techniques are usually static, offering no possibility for post-fabrication adaptation.

Tunable polariton optics can be realized by combining $\alpha$-MoO$_3$ with tunable or programmable substrates since the polariton propagation can be tailored by substrate engineering (c.f. Supplementary Note 2 for comparative literature study). Global tuning of polariton wavelength and direction have been realized by placing α-MoO$_3$ flakes on different substrates[20] or tailoring the optical properties of the substrate[12,16,19]. However, optical structures such as a focusing structures for these tunable polaritons still rely on conventional time-consuming fabrication techniques. Fast prototyping of polariton optics instead has been realized by locally changing the optical conductivity of the oxide SmNiO$_3$ and afterwards aligning an α-MoO$_3$ flake.[25] While polariton optics have been directly imprinted, this approach has not been used for prototyping after flake deposition nor for reconfiguring polariton optics.

Fast prototyping of reconfigurable polariton optics can be realized with phase-change materials (PCMs). PCMs can reversibly be switched between two (meta-)stable phases, an amorphous and a crystalline phase, by inducing heat, e.g. by optical laser pulses.[26,27] The two phases do usually feature distinct electronic and optical properties. For example, the two dielectric phases of GeSbTe (GST) alloys feature a permittivity contrast of about 24 in the infrared range between 800 and 970 cm$^{-1}$.[26] The pronounced contrast of the optical properties in PCMs like Ge$_2$Sb$_2$Te$_5$ has been attributed to a change of bonding between the amorphous and crystalline state[27,28]. While amorphous PCMs employ ordinary covalent bonding, crystalline Ge$_2$Sb$_2$Te$_5$ utilizes a special bonding mechanism called metavalent bonding[29]. This bonding mechanism is characterized by unconventional properties[30] and an unusual bond rupture in atom probe tomography[31]. This permittivity contrast has been exploited to refract hyperbolic phonon polaritons in hBN launched at a flake edge by an optically programmed lens in the GST below.[32] Recently, the PCM In$_3$SbTe$_2$ (IST) has been introduced for fast prototyping of reconfigurable nanophotonics[33] such as antennas[34–38] and advanced metasurfaces[39,40] for emissivity control, beam-steering, lensing and holography. The PCM IST stands out as it can be switched between a dielectric amorphous ($\epsilon_{aIST}$ = 14 > 0) and a metallic crystalline phase ($\epsilon_{cIST}$ = -159 < 0) featuring a permittivity contrast of about 173 in the infrared range between 800 and 970 cm$^{-1}$[34]. Thus, the PCM IST is a promising platform for fast optical prototyping of reconfigurable launching structures for polaritons in the infrared[41–43]. IST has been used to launch and confine surface phonon polaritons in SiC[41], surface plasmon polaritons in the doped semiconductor CdO[42], and hyperbolic phonon polaritons in hBN[43,44]. Since most vdW materials are highly transparent in the visible spectral range that is used for optical programming of structures in IST, the structures can be programmed through the already deposited flake[43]. Importantly, this optical programming after flake deposition allows for alignment of the structures to the flake axes, for reconfiguration of previously written structures, and for addition of structures to the very same flake. While the modification of the α-MoO$_3$ polariton dispersion by IST as substrate and optical programming of a focusing structure have already been demonstrated[44], the major advantage of programming polariton optics after flake deposition and their reconfiguration have not been demonstrated yet. Here, we deposit the in-plane hyperbolic vdW material α-MoO$_3$ onto the phase-change material IST. After flake deposition, we rapidly prototype polariton optics through the α-MoO$_3$ flake into the IST within minutes (see Supplementary Note 1). First, we retrieve the experimental dispersion of polaritons launched at a phase boundary of IST below the α-MoO$_3$ flake performing sequential near-field optical spectroscopy. We optically program stripes of arbitrary angle with respect to the crystallographic axis of α-MoO$_3$ to steer the high directionality of polaritons in α-MoO$_3$. By optically programming a circular crystalline disk into the IST, we achieve focusing of α-MoO$_3$ polaritons that can be tailored by varying the illumination frequency. Finally, we create a nanocavity for the in-plane hyperbolic polaritons by adding a second disk in a post-processing step. To further tailor the polariton confinement, we vary the disk distance. Our experimental findings, supported by numerical simulations, demonstrate the promising potential that IST holds for real-time, on-demand (re-)programming of polariton optics for vdW materials and for impacting the next generation of compact nanophotonic devices.

**Results**

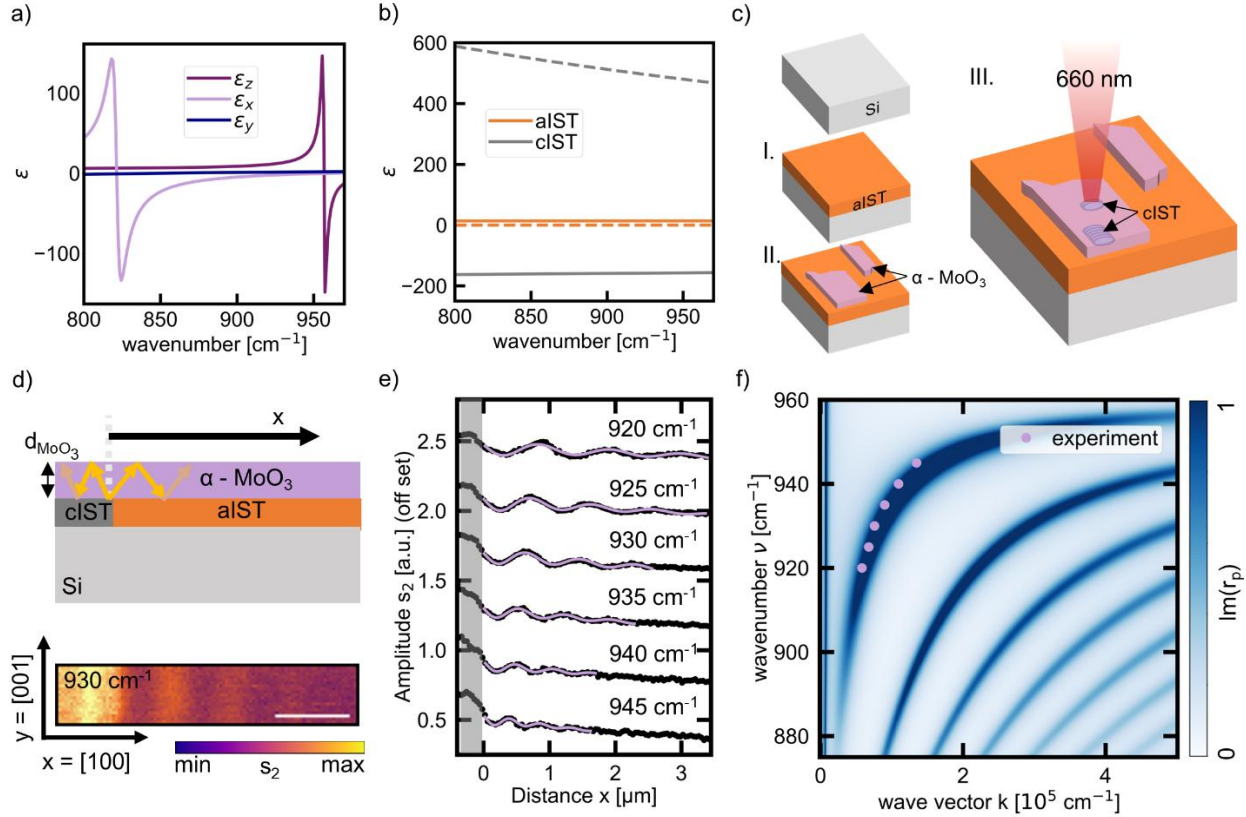

**Figure 1: Direct programming of α - MoO₃ phonon polariton launching structures into In₃SbTe₂ (IST).** *a) Real part of the permittivity ε of α-MoO₃ and b) real part (solid) and imaginary part (dashed) of the permittivity ε of IST in the reststrahlenband (RB2) of α-MoO₃ where permittivity along the optical x-axis εₓ is negative. c) Schematic layer stack fabrication and subsequent programming of optical structures. d) Cross sectional sketch of the investigated structure programmed into the IST by local optical crystallization to launch phonon polaritons in the α-MoO₃ (d_{α-MoO₃} = 392 nm) above (top). Exemplary s-SNOM amplitude image of the phonon polaritons launched at the boundary of amorphous and crystalline IST (bottom), extracted from the same images that will be used in Figure 3. e) Line profiles extracted from s-SNOM amplitude images at a boundary of amorphous and crystalline IST as in d) for different illumination frequencies fitted assuming only edge-launched polaritons (purple lines). f) Experimental (purple dots, extracted from the line profiles in e)) and simulated dispersion (colorplot of the imaginary part of the Fresnel reflection coefficient) of the phonon polaritons of α-MoO₃ on top of amorphous IST.*

Aiming to steer the anisotropic polaritons in α-MoO₃, we combine the vdW material flakes with the

PCM IST that allows for optical programming of reconfigurable polariton launching optics. α-MoO₃

is a highly anisotropic vdW material that has distinct permittivities along all three crystal axes[45]. In the following, we choose the coordinate system such that the x-axis is aligned with the [100], the y-axis with the [001] and the z-axis with the [010] direction of the α-MoO₃ crystal. Remarkably, the phonon resonances along the optical x-, y- and z-axis of the flake occur at different frequencies such that the reststrahlenbands (RBs) – the spectral range between the transverse and the longitudinal optical phonon resonances, where the permittivity is negative – do barely overlap. We will focus on the RB that spans from 821.4 to 963 cm$^{-1}$ (commonly referred to as RB2) where the permittivity along the optical x-axis is negative while in most of the spectral range the permittivity along optical y- and z-axes is positive[45] as depicted in Figure 1a. Consequently, the isofrequency curve (IFC) in the corresponding Fourier space ($k_x$,$k_y$) is hyperbolic[46] (see Supplementary Note 3). This leads to strongly anisotropic propagation of polaritons in α-MoO$_3$[5,6,10]. As stated above, IST can be switched between the dielectric amorphous phase (aIST) characterized by a positive permittivity and the metallic crystalline phase (cIST) characterized by a negative permittivity (c.f. Figure 1b). Hence, IST serves as a substrate that can be locally switched between a dielectric and metallic phase and thus offers fast local programming of reconfigurable polariton optics. Thereby, we use a fabrication scheme that overcomes the drawbacks of conventional multistep fabrication processes for polariton optics (c.f. Supplementary Note 1). Our fabrication scheme is based only on three steps (c.f. Figure 1c). First, the IST film is sputtered in the amorphous phase onto a silicon wafer, followed by sputtering of a capping layer of ZnS:SiO$_2$ to prevent the IST from oxidation (omitted in sketches). Secondly, α-MoO$_3$ is mechanically exfoliated and transferred (see Methods) onto the IST thin film resulting in several α-MoO$_3$ flakes of different thicknesses and in-plane dimensions deterministically distributed over the sample. Thirdly, we optically program the desired structure through the α-MoO$_3$ flake into the IST by locally crystallizing the IST using a pulsed laser with a wavelength of 660 nm (see Methods). By spatially overlapping several laser pulses, we can program arbitrary shapes of cIST at any position as sketched in Figure 1c. Compared to conventionally used fabrication schemes we identify three major advantages: 1) The fewer fabrication steps lead to much faster turn-around times. Using our fabrication scheme launching structures can be created within a few hours. 2) We can optically program structures after flake deposition. Thus, we can use the much faster deterministic flake transfer and can align the structure to the flake axes while in conventional approaches the flake needs to be aligned with the structure. 3) Our fabrication allows for post-processing adaptation. We can reconfigure the already written structure and can adapt size and shape by further crystallization. Further, we can add additional structures to the same flake (or other flakes on the same wafer) within 10 minutes as we do not need to repeat the first two steps.

We first explore polaritons launched at a phase boundary in the IST below the α-MoO₃ as sketched in Figure 1d top. Due to the vast permittivity difference between amorphous and crystalline IST, polaritons are launched at the phase boundary. The hyperbolic polaritons of α-MoO₃ are volume-confined, i.e. the electric field of the polaritonic mode is confined within the thin α-MoO₃ slab[47], and propagate through α-MoO₃ towards both sides of the boundary. We employ scattering-type scanning near-field optical microscopy (s-SNOM)[48–50] (see Methods) to image the phase-boundary-launched polaritons that interfere with the incident light similar to previously investigated phase-boundary-launched polaritons in hBN[43]. We do not observe tip-launched and edge-reflected but only phase-boundary-launched polaritons that interfere with the incident light similar to edge-launched polaritons[51–53]. An exemplary s-SNOM amplitude image at 900 cm⁻¹ of a phase boundary below the α-MoO₃ is shown in Figure 1d bottom. This image depicts an area of high s-SNOM amplitude on the left and a lower s-SNOM amplitude on the right corresponding to the metallic cIST with highly negative permittivity on the left and the dielectric aIST with a positive permittivity on the right, respectively. Within the area of aIST we observe bright fringes parallel to the phase boundary. These bright fringes correspond to polaritons launched at the phase boundary and propagating in the α-MoO₃ above aIST. To quantify the polariton dispersion we extract line profiles perpendicular to the phase boundary in the area of aIST from s-SNOM amplitude images like that in the Figure 1d bottom for different frequencies (Figure 1e). From the line profiles we extract the polariton wavevector $k_p$ (see Supplementary Note 4 for details) that corresponds to the polariton wavelength $\lambda_p$ via $k_p = \frac{2\pi}{\lambda_p}$ to determine the experimental dispersion (purple dots in Figure 1f). With increasing frequency, the polariton wavevector (wavelength) increases (decreases). The experimentally observed dispersion agrees with the theoretical prediction from the imaginary part of the Fresnel reflection coefficient (blue branches in Figure 1f) that we calculated using the transfer matrix method[54].

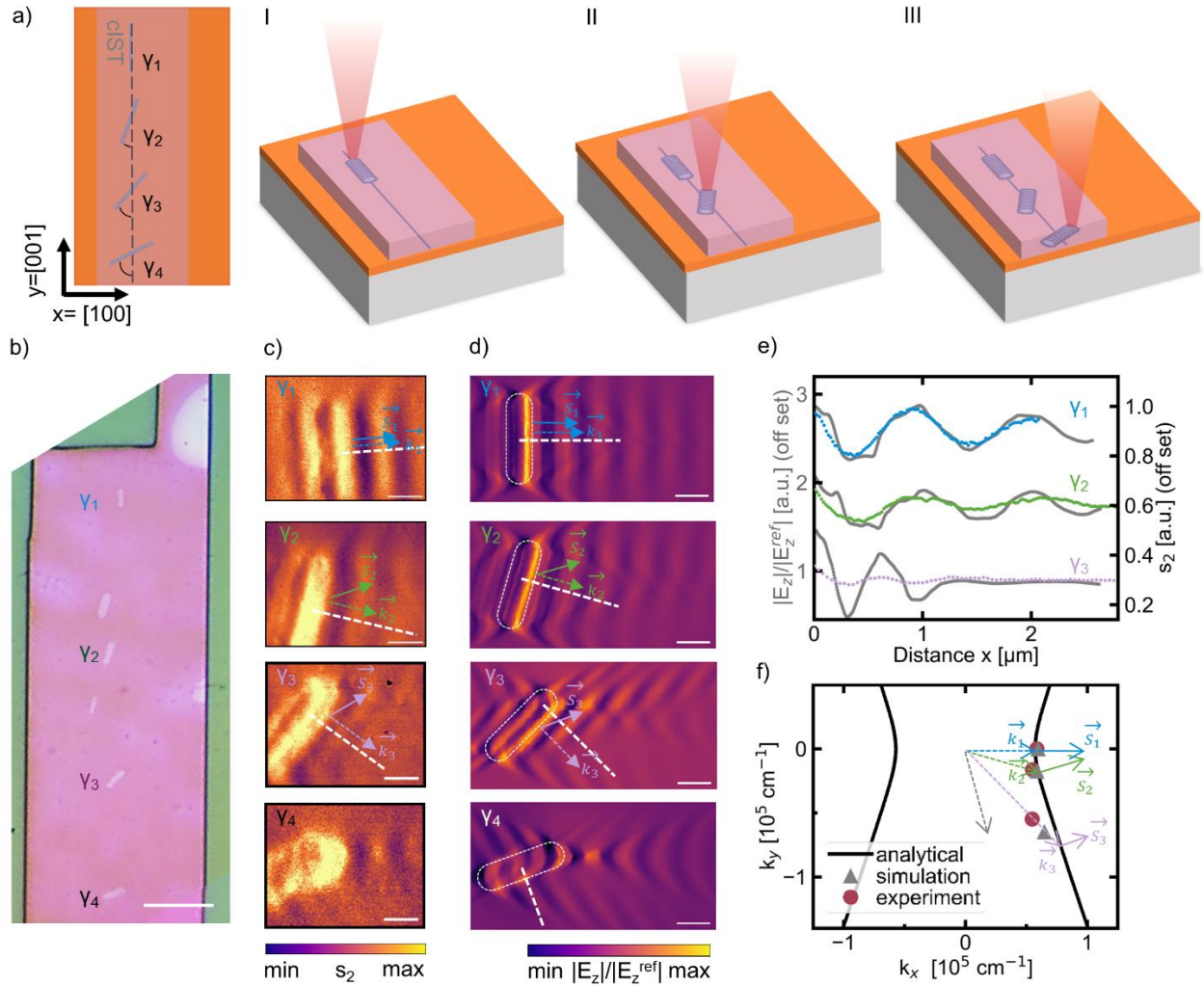

**Figure 2: Fast prototyping of launching stripes of different angles with respect to the crystal axis:** a) Design sketch and optical writing process of stripes tilted by different angles with respect to the optical y-axis of the crystal. b) Optical microscope image of the tilted antennas programmed into the IST below the α-MoO₃ flakes ($d_{α-MoO_3}$ = 252 nm). The scale bar corresponds to 5 µm. c) s-SNOM amplitude images at 900 cm⁻¹ and d) corresponding simulation of antennas with four different tilt angles. The scale bars correspond to 1 µm. In b)-d), the angle between the stripe long axis and the optical y-axis of the flake increases from top to bottom with $γ_1$=-2.4°, $γ_2$=16.5°, $γ_3$=44.7°, and $γ_4$=70.2°. The wavefronts are parallel to the long axes of the stripes and thus the wavevector are perpendicular to the long axes of the stripes (indicated by the dashed arrows). In contrast, the direction of energy flow (Poynting vector: solid arrows) varies depending on the tilt angle. e) Line profiles perpendicular to the stripes long axes extracted from the s-SNOM amplitude images (color corresponding to the tilt angle as in b)-d)) and from simulations (grey) as indicated by the white dashed lines in c) and d), respectively. f) IFC with wavevectors extracted from experiment (red dots) and simulation (grey triangles). The dashed arrows indicate the wavevector and the solid arrows the Poynting vectors.

Exploiting our fast fabrication scheme that also allows for (re-)programming of launching structures through already deposited flakes and thus for their flexible alignment with the optical axes of the flake (Figure 2a), we study the in-plane anisotropic propagation of the polaritons. To investigate the directional propagation, we optically program crystalline launching stripes with different orientations with respect to the optical y-axis of the flake as sketched in Figure 2a. The concept of launching stripes with different orientations is similar to those conducted in literature using conventionally fabricated metallic launching stripes[12] and etched trenches[10]. The angle γ between the long axis of the stripes and the y-axis of the flake is increased sequentially from top to bottom. In the optical light microscope image (Figure 2b right), the crystalline stripes appear as bright areas within the pink α-MoO$_3$ flake. We investigate the anisotropic polaritons in α-MoO$_3$[5,6,10] in the RB2 where the permittivity along the optical x-axis is negative. Thus, from the IFC which graphically displays the solution of the dispersion relation at a fixed frequency in Fourier space (c.f. Supplementary Note 3 and Supplementary Note 9), we expect the propagation to be preferential along the x-direction and forbidden along the y-direction, as observed in literature[5,6,10]. The propagation direction is restricted to a cone defined by the complementary of the opening angle of the IFC. Experimentally, we use the s-SNOM to visualize the propagating polaritons (shown in Figure 2c) and support our experimental results with simulations of the electric field (see Figure 2d).

In Figure 2c top image, the stripe is almost aligned with the y-axis, thus $\gamma_1 \sim 0°$. We observe bright fringes parallel to the long axis of the stripe (y-axis) and periodic in x-direction. These bright fringes originate from polaritons launched by the cIST stripe and interfering with the incident light. The direction of polariton propagation (perpendicular to the bright fringes; x-direction) corresponds to the direction of the wavevector (indicated by the blue dashed arrow) and is perpendicular to the stripe. Also, the direction of energy flow corresponding to the direction of the Poynting vector (indicated by the blue solid arrow) is perpendicular to the stripe. In this case, the Poynting vector and wavevector are collinear. To further analyze the polariton propagation, we extract line profiles perpendicular to the stripe from both s-SNOM amplitude and simulation images. The resulting line profiles are shown in Figure 2e. By fitting the line profiles (see Supplementary Note 4), we experimentally determine the polariton wavelength, and, thus, the magnitude of the polariton wavevector. Figure 2f displays a direct comparison between the experimentally determined wavevector (red dots) and the simulation (grey triangles), for each angle γ. The black solid line represents the analytically calculated IFC of the system. For $γ_1$~0° the wavevector is parallel to the $k_x$-axis and $k_y = 0$. Considering the IFC, we see that the Poynting vector whose direction is always perpendicular to the IFC for $γ_1$~0° is collinear with the wavevector and also along the x-direction in good agreement with our experimental observations.

When increasing the angle γ, we observe, both in the experimental (Figure 2c) and in the simulated (Figure 2d) images, again bright fringes parallel to the stripe for $γ_2$ and $γ_3$ whereas for $γ_4$ no fringes parallel to the long axis of the stripe are observed. For increasing angle γ between the long axis of the stripes and the y-axis of the flake in the s-SNOM amplitude images in Figure 2c and simulations in Figure 2d, we observe again bright fringes parallel to the stripe for $γ_2$ and $γ_3$ whereas for $γ_4$ no fringes parallel to the long axis of the stripe are observed. For increasing angle γ, the spacing between the fringes decreases. While the wavevector of the polaritons (indicated by dashed arrows) launched by the stripes with $γ_2$ and $γ_3$ is perpendicular to the stripes as it is for $γ_1$~0, the direction of energy flow (indicated by solid arrows) is no longer perpendicular to the stripe nor collinear with the direction of propagation. For a qualitative understanding, we again extract line profiles (Figure 2e), determine the polariton wavelength and calculate the magnitude of the polariton wavevector. In agreement with the observation of reduced spacing of the fringes, we find a reduced polariton wavelength for an increased angle γ corresponding to an increased polariton wavevector. To plot the wavevector into the IFC, the angle of propagation with respect to the optical x-axis has to be considered that is given by the angle γ between the long axis of the stripes and the y-axis of the flake. As for a non-zero angle γ the wavevector is no longer aligned with the x-axis, the wavevector component $k_y$ is non-zero. The wavevectors extracted from simulation and experiment agree well and follow the trend of the theoretical IFC. As conducted in literature[10,12,22,55] by plotting the wavevector into the IFC and considering the propagation direction the non-collinearity of wavevector and Poynting vector can be explained. Since the Poynting vector is normal to the IFC, for any wavevector direction not aligned with the optical x-axis, the Poynting vector is not collinear with the wavevector. Direction of energy flow and propagation direction are not collinear explaining the observations in the s-SNOM images and simulations.

If the angle γ between the long axis of the stripes and the y-axis of the flake exceeds a the opening angle of the IFC the wavevector direction would be such that no intersection with the IFC is possible (see grey dashed arrow in Figure 2f) and thus propagation along this direction is not allowed.[10,12,22] We do not observe fringes parallel to the stripe aligned with angle $\gamma_4$ (c.f. Figure 2c) as the angle $\gamma_4$ exceeds the opening angle of the IFC.

Relying on optically programmed launching stripes in IST we can retrieve the IFC experimentally, thereby characterizing the directionality and at the same time steer the direction of polariton propagation similar to previous experiments relying on conventionally fabricated metallic launching stripes[10,12]. In contrast to conventional fabrication techniques, we can speed up the fabrication process as we drastically decrease the turn-around times and can flexibly align the structures to the flake axes facilitating characterization and steering of anisotropic polaritons.

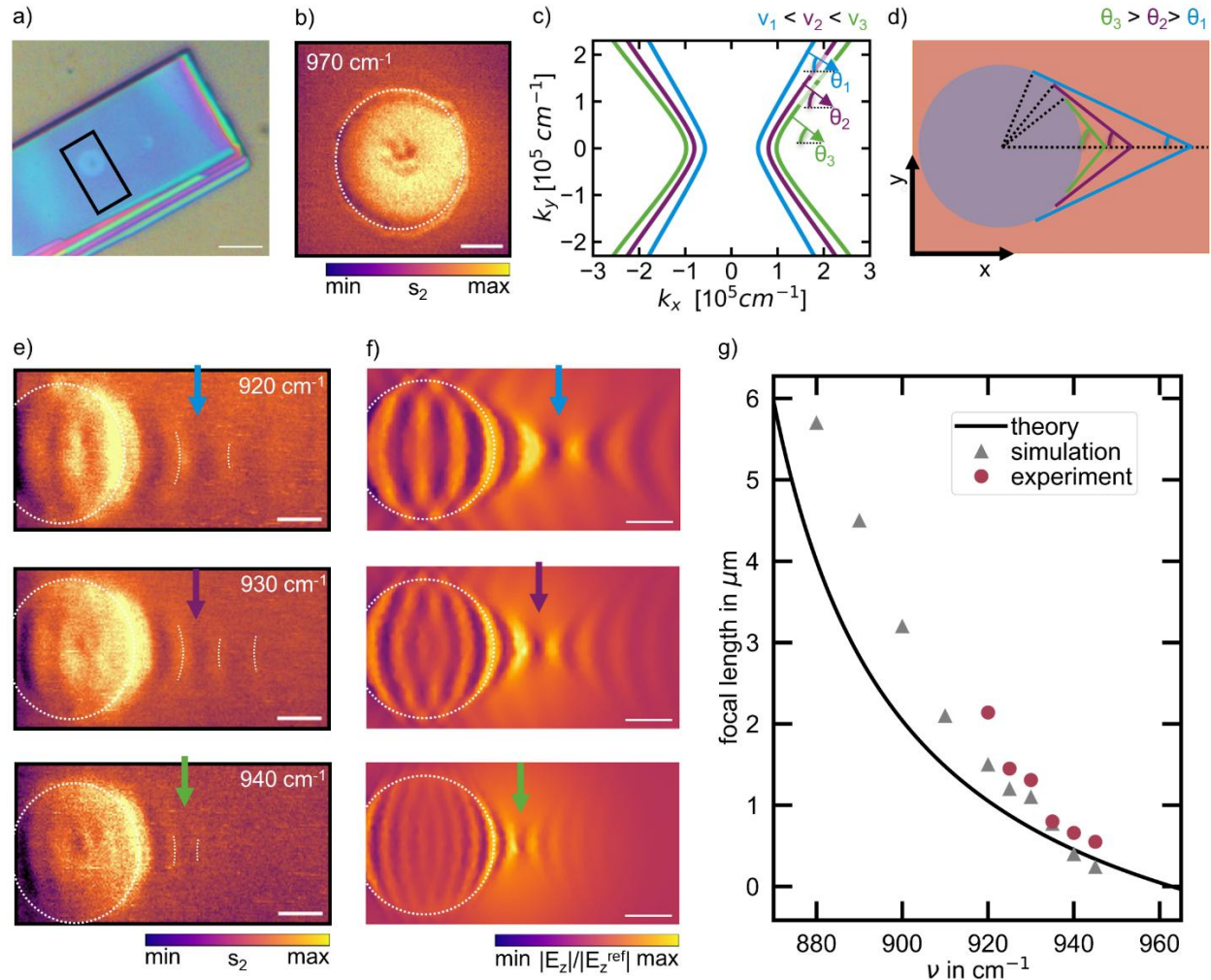

***Figure 3: Optically programmed focusing structure.*** *a) Optical microscope image of an optically programmed circular crystalline launching structure in IST (cIST disk) below α-MoO₃ ($d_{α-MoO_3}$ = 392 nm). b) s-SNOM amplitude image of the circular crystalline launching structure in RB3 of α-MoO₃ at 970 cm⁻¹. At this frequency the polaritons propagate under an angle close to zero through the α-MoO₃ providing a direct image, also called super-resolution image, of the buried cIST disk. c) IFC for three different frequencies increasing from 920 (blue) to 930 (purple) to 940 cm⁻¹(green). d) Sketch of the focusing of cIST disk below the α-MoO₃ for the same frequencies as in c). e) s-SNOM amplitude images and f) electric field simulations of the focusing of α-MoO₃'s phonon polaritons by the cIST disk for three exemplary frequencies (same as in c) and d)). The scale bars correspond to 1 μm. g) Focal length as a function of incident frequency extracted from s-SNOM amplitude images (purple dots) and simulations (grey triangles) as in e) and f) together with the theoretical behavior (black solid line).*

We now optically programm a circular cIST launching structure (cIST disk) to focus the polaritons.

In the optical image in Figure 3a the cIST disk is visible as a bright circle. As we program the cIST disk after having deposited α-MoO₃, we cannot characterize its exact size and shape by direct measures (e.g. by atomic force microscopy as it is done for conventionally fabricated structures).

We overcome this issue by performing a super-resolution s-SNOM amplitude image (Figure 3b) at 970 cm$^{-1}$, close to the lower bound of the RB3. At this frequency, the polaritons propagate at an angle close to zero through the slab such that they provide a direct image of the buried structure, a so-called super-resolution image[56–61] (see Supplementary Note 5). Therefore, we interpret the circular structure of high amplitude as a direct image of the cIST disk below the α-MoO$_3$ flake.

Due to the in-plane hyperbolicity of α-MoO$_3$ in the RB2, the cIST disk is expected to launch polaritons at any point of its circumference which are focused to a common point along the x-axis as sketched in Figure 3d as observed previously for gold launching disks[17,19,20], gold antennas[18] and bent silver nanowires[21] on top of α-MoO$_3$. The focal length is determined by the radius of the launching structure and the the complementary angle $\theta = \frac{\pi}{2} - \alpha$ to the opening angle $\alpha$ of the IFC[19]

$$: f = R\left(\sqrt{1 + \frac{1}{\tan^2\theta}} - 1\right) = R\left(\sqrt{1 - \frac{\varepsilon_x(\nu)}{\varepsilon_y(\nu)}} - 1\right).$$

The second step uses the relation for the opening angle of the IFC $\tan(\alpha) = \sqrt{-\frac{\varepsilon_x(\nu)}{\varepsilon_y(\nu)}}$, which is valid for thin α-MoO$_3$ flakes. Since the opening angle of the IFC decreases with increasing illumination frequency (c.f. Figure 3c), the focal length decreases with increasing illumination frequency, as also reported experimentally[17–19,21].

To visualize the focusing of the phonon polaritons in α-MoO$_3$ by the cIST disk and to study the tuning of the focal length experimentally, we show s-SNOM amplitude images at three different frequencies in the RB2 in Figure 3e. We support our s-SNOM amplitude images with electric field simulations (c.f. Figure 3f). In both experiment and simulation, we observe curved bright fringes (indicated by white dashed lines) right to the cIST disk (marked by white dashed circumference) that first resemble the curvature of the cIST disk and can be described with a narrowing envelope. We identify the focal point where the curvature of the wave fronts is inverted and the envelope diverges. We marked the focal point with arrows in experiment and simulation images (c.f. Figure 3e and 3f). In agreement with the theoretical expectations, with increasing illumination frequency the focal length decreases and the fringe spacing corresponding to the polariton wavelength decreases. From s-SNOM amplitude images and simulations such as shown in Figure 3e and 3f, we extract the focal point as a function of illumination frequency (c.f. Figure 3g). The experimentally extracted focal point agrees well with that found in our simulation study. However, especially for lower frequencies (longer focal lengths) the focal length extracted from experiment and simulation deviate from the theoretical expectation which relies on the assumption of thin flakes. A similar deviation has been observed in literature when $\alpha$-MoO$_3$ flake thicknesses are above 300 nm.[19] Since our flake is 392 nm thick, our findings agree with that in literature. Our results on phonon polaritons in $\alpha$-MoO$_3$ focused by an optically programmed cIST disk agree well with that observed by conventionally fabricated metallic focusing structures in literature[17–21]. Tuning the illumination frequency, we demonstrate shifting the focal length from 0.55 to 2.14 µm. Our achieved tuning range of the focal length is slightly larger than that achieved by Martín-Sánchez et al.[17] (0.6-1.7 µm) and Zheng et al.[18] (0.6-1.6 µm) who relied on conventionally fabricated focusing structures. Thereby, we demonstrate that our fast optical programming of cIST disks allows for similar focusing qualities as conventionally fabricated metallic disks while offering much faster turn-around times in fabrication and the opportunity for post-processing adaptation.

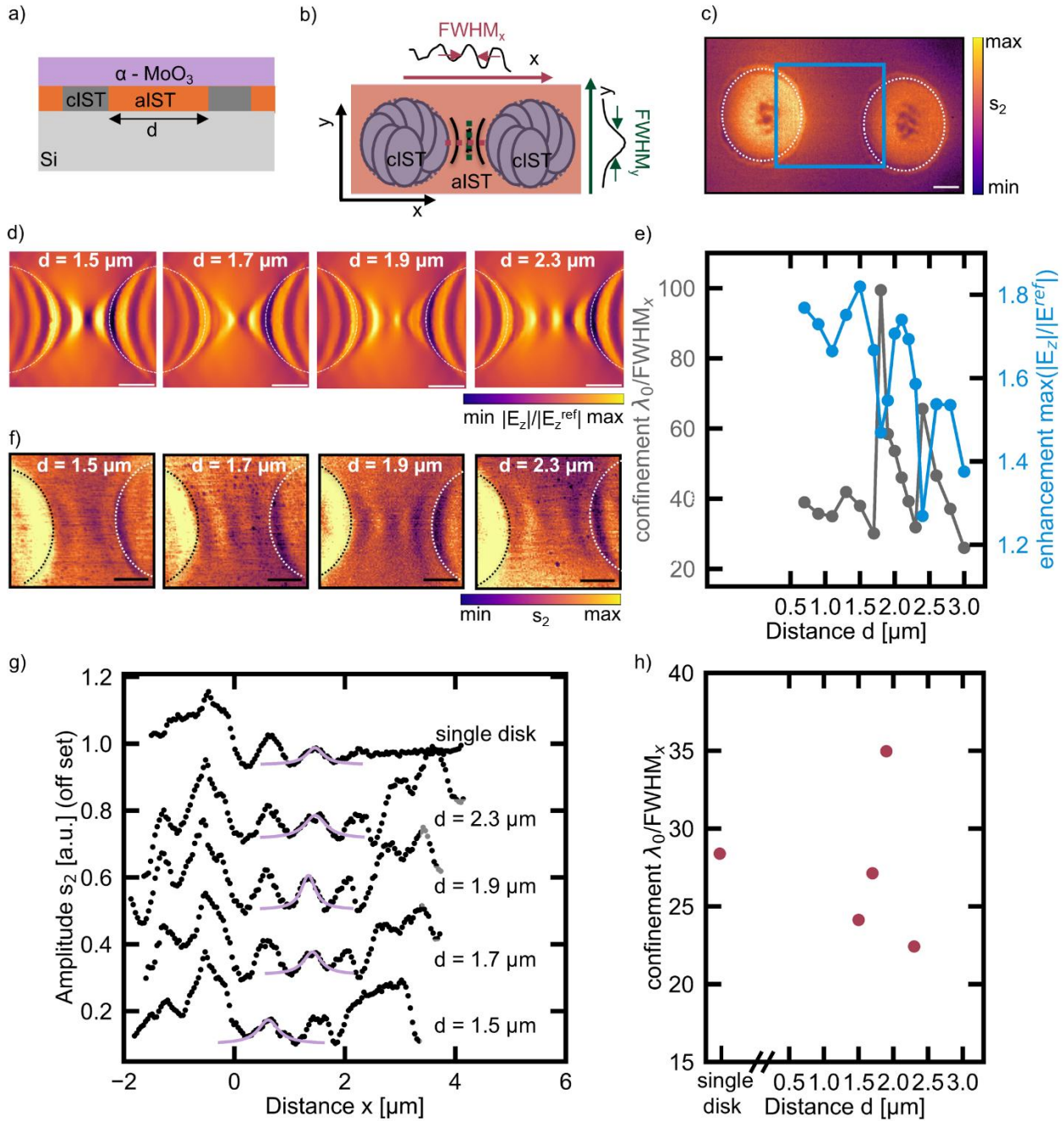

**Figure 4: Matching the focal point of two optically programmed focusing structures to confine phonon polaritons in α-MoO₃.** Sketched a) side view ($d_{α-MoO_3}$ = 392 nm) and b) top view of two crystalline launching structures whose focal points overlap. c) Super-resolution image of the optically programmed double disks taken at 970 cm⁻¹ with s-SNOM. d) Simulations and f) s-SNOM amplitude images at 930 cm⁻¹ of the two focusing disks for four different distances increasing from left to right. The scale bars correspond to 1 μm. e) Field confinement $λ_0/FWHM_x$ (c.f. inset in b)) and enhancement along the optical x-axis at the focal point as function of disk distance d extracted from simulations such as displayed in d). g) Line profiles of a single launching disk (top) extracted from s-SNOM images in Figure 3e) and line profiles of two launching disks whose distance decreases from top to bottom extracted from s-SNOM images in f). h)

*Confinement $\lambda_0/FWHM_x$ along x-direction determined from line profiles extracted from experimental s-SNOM images (red dots) as a function of disk distance.*

We aim for further tailoring both the confinement and enhancement of polaritons. In a post-processing step, we add a second cIST disk to the focusing cIST disk discussed in Figure 3 (see sketch in Figure 4a and 4b). By spatially overlapping the two focal points, we create a nanocavity for the in-plane hyperbolic polariton. To our best knowledge, this approach has not been followed experimentally in previous literature. Our geometry was inspired by the theoretically proposed hyperbolic nanocavity as nanocavity geometry for in-plane hyperbolic polaritons as a counterpart to circular nanocavitys for in-plane isotropic polaritons by Álvarez-Pérez et al.[22]. For the double disk structure, we expect enhanced confinement and field enhancement compared to a single disk. The confinement along x- and y-direction is determined by the full-width-at-half maximum ($FWHM_{x/y}$) in x- and y-direction of the maximum closest to the middle of the two disks as sketched in Figure 4b in red and green, respectively. Figure 4c shows a super-resolution s-SNOM image taken at 970 cm$^{-1}$ of the two crystalline disks in the amorphous surrounding. We use this super-resolution image to determine the distance between the disks experimentally to be 1.9 µm.

To prove tailoring the confinement and field enhancement of polaritons, we first perform numerical simulations to tailor the interference of the polaritons by varying the disk distance d. Figure 4d shows simulations of the electric field along z-direction for four exemplary disk distances increasing from left to right from 1.5 µm to 2.3 µm. The displayed simulations show the area between the two cIST disks marked by the blue rectangle in Figure 4c. Between the two cIST disks that are marked by white dashed circumferences, we observe alternating minima and maxima along the x-direction. By increasing the disk distance, the number of maxima between the two disks increases and the FWHM of the maxima in x- and y-direction are tailored. As the FWHM is directly related to the confinement of the field $\lambda_0/FWHM_{x/y}$ in x- and y-direction, respectively, we expect a strong dependence of the confinement on disk distance. To quantify this, we extract line profiles (as sketched by the dashed lines in Figure 4b) from the simulations along x-(red) and y-direction (green). We determine the FWHM of the maximum closest to the middle of the two disks as indicated in the line profiles inset and sketched in Figure 4b to calculate the confinement. Further, we determine the field enhancement (max. value of the referenced electric field) from the same maximum. Confinement and field enhancement along the x-direction from simulation are displayed as function of disk distance in Figure 4e. The field enhancement along the x-direction from simulation is maximal at disk distances of about 1.5 (field enhancement of 1.8) and 2.1 µm (field enhancement of 1.7). These distances correspond to the cases where the polaritons launched by the two disks interfere constructively. The maxima close to the center overlap leading to a maximal field enhancement (detailed discussion in Supplementary Note 6). The confinement along the x-direction from simulation instead is maximal at 1.8 µm disk distance ($\lambda_0/99$), close to the geometric overlap of the focal points of the two disks. The maximal confinement along the y-direction of $\lambda_0/45$ we find at a disk distance of 2.4 µm (see analysis in Supplementary Note 7). Consequently, analyzing the simulations we find that by tailoring the disk distance we can either maximize field enhancement or confinement.

Exploiting the fast prototyping of our optically programmed double disks, we further corroborate our numerical findings by studying the polariton interference for four different distances by taking s-SNOM amplitude images (see Figure 4f). The distances correspond to the distances for which we show simulations in Figure 4d. In the s-SNOM amplitude images we observe similar patterns as in the simulations. To quantify the confinement in x-direction, we extract line profiles in x-direction through the middle of the two disks (c.f. red dashed line in sketch in Figure 4b) and compare them to a line profile extracted for a single launching disk in Figure 4g. We determine the $FWHM_x$ of the maximum closest to the middle (for the single disk closest to the focal point) by fitting a Lorentzian function. The experimental confinement in x-direction is plotted in Figure 4h. We observe an increased confinement along the x-direction of $\lambda_0/35 = \lambda_P/2.7$ at a disk distance of 1.9 µm in good agreement with the observed maximal confinement in simulations at 1.8 µm. This disk distance corresponds to the distance where the focal points of left and right disk approximately overlap. Along the y-direction (see analysis in Supplementary Note 7), we find an enhanced confinement of $\lambda_0/20 = \lambda_p/1.4$ at a disk distance of 1.7 µm. Compared to the confinement of the single disk ($\lambda_0/28 = \lambda_p/2.1$ along x-direction and $\lambda_0/10 = \lambda_p/0.8$ along y-direction), the double disk arrangement allows for an increase in the confinement along x-direction by a factor of 1.25 and along y-direction by a factor of 2. Our results on confinement for the single disk arrangement are comparable to that observed in literature[12,17–20], e.g. Martín-Sánchez et al.[17] investigated a single gold disk and achieved a confinement of $\lambda_0/28$ along the x-direction and of $\lambda_0/34$ along the y-direction (c.f. detailed comparative literature study in Supplementary Note 2). Further tailoring of confinement can be achieved by tuning the illumination frequency (see Supplementary Note 8). We envision confinement and field enhancement could be further enhanced by tailoring size and distance of the two disks. Moreover, the interference and thus confinement and field enhancement might depend on the launching phases.[22] Thus, investigating the launching phase and its influence on confinement and enhancement by different shapes of the launching structure and different illumination directions is of high interest for future works. Field enhancement and confinement might be even further enhanced when adapting the shape of the launching structures such that they resemble the complementary of the IFC[17,22] and using thinner flakes as the confinement generally increases for thinner flakes.

**Discussion**

In this work, we combine the polariton-hosting vdW material α-MoO$_3$ with the phase-change material IST as this allows for fast prototyping of reconfigurable polariton optics. Polariton launching structures can optically be programmed into the IST through the α-MoO$_3$ flake. Our fabrication scheme stands out due to the few fabrication steps needed and thus fast turn-around times, due to the optical programming of structures after flake deposition and thus flexible alignment of the structure with the flake axes, and due to the opportunity for reconfiguration and addition of structures to already fabricated structures on the very same flake. Hence, it is a versatile platform to steer and study anisotropic polaritons fast and efficiently. We optically program launching stripes through the exfoliated α-MoO$_3$ flake what allows us to align the launching structures of different orientations with the crystal axis. The aligned launching stripes enable steering the propagation direction of the polaritons and also retrieving the IFC experimentally. We exploit the fast optical programming of crystalline launching disks to focus the polaritons. By varying the illumination frequency, we tailor the focal length from 0.55 to 2.14 µm. At 930 cm$^{-1}$ we find a field confinement of $\lambda_0/28 = \lambda_p/2.1$ along x-direction and $\lambda_0/10 = \lambda_p/0.8$ along y-direction. Both the tuning range of the focal length and the confinement achieved by an optically programmed cIST disk are comparable to values in literature for conventionally fabricated metal disks demonstrating the high quality of the optically programmed structures. Finally, we add a second disk to the first focusing disk in a post-processing step. Thereby, we fabricate a nanocavity for the anisotropic polaritons in α-MoO$_3$ which has by now only been proposed theoretically[22]. We add double disks of different disk distances to the very same flake to tailor field confinement and enhancement. We find a maximal field confinement along the x-direction of $\lambda_0/35 = \lambda_p/2.7$ at disk distance of 1.9 µm and along the y-direction of $\lambda_0/20 = \lambda_p/1.4$ at a disk distance of 1.7 µm. By employing the double disk structure, we can enhance the confinement along the x-direction by a factor of 1.25 and along y-direction by factor of 2 compared to the single disk.

The programming scheme is not limited to single nanostructures but in-principle much more complex structures such as metasurfaces could be optically programmed into the IST similar to metasurfaces programed into GST to focus polaritons in hBN[32]. Employing IST as substrate for optical programming, consequently, is promising for fast prototyping of various complex polariton optics.

Optical programming of polariton optics for vdW materials with IST facilitates and speeds up the fabrication of launching structures that enable steering of polaritons. As steering of polaritons is essential for nanophotonic devices, the fast prototyping of reconfigurable polariton optics with IST might facilitate further research on nanophotonic devices. Further, launching structures are needed to characterize the directionally of polaritons in vdW materials. Characterization of highly directional polaritons in other vdW materials such as the in-plane anisotropic vdW materials β-$Ga_2O_3$[62], α-$V_2O_5$[8] or $LiV_2O_5$[9] can be facilitated by the fast prototyping of polariton optics in IST. We envision the optical programming of polariton optics for vdW materials with IST can also facilitate and fasten the exploration of fundamental optical phenomena on the nanoscale such as high directionality[10–12], planar refraction[13,14], focusing[14,17–21] and negative reflection[22] and also topological transitions that have been studied with coupled polaritons in vdW heterostructures[12].

**Methods**

**Sample Fabrication.** Using direct current and radio frequency sputter deposition, a 50 nm (sample 1 in Figure 1, Figure 3 and Figure 4) and 100 nm (sample 2 in Figure 2) thick layer of amorphous $In_3SbTe_2$ and a 20 nm thick capping layer of $ZnS:SiO_2$ were applied on a Si substrate. The capping layer (80% ZnS, 20% $SiO_2$) prevents the sample from oxidation. α-$MoO_3$ thin slabs were prepared by mechanical exfoliation of bulk α-$MoO_3$ crystals (Alfa Aesar) using Nitto adhesive tape (Nitto Denko Co., SPV 224P). To further reduce the thickness, a second exfoliation step was performed by transferring the single thin slabs from the tape onto a transparent polydimethylsiloxane (PDMS) stamp. The resulting thin slabs were examined under an optical microscope, and homogeneous regions with large lateral dimensions were selected. These single thin slabs were then deterministically transferred onto the previously described amorphous PCM sample using a dry-transfer technique[63].

**Optical switching.** A home-built laser switching setup was used to crystallize the PCM locally. Here, the light of a 660 nm laser diode is focused on the sample with a 10-fold objective with NA = 0.25. Each individual crystallized spot for the disk structures was created by applying 100 pulses with a power of 110 mW and a pulse length of 800 ns. Due to the elliptically beam profiles the crystallized spots are elliptical. One elliptical crystallized spot has a lateral extension of approximately 1 x 1.5 $\mu m^2$. For crystallizing the crystalline launching stripes, 21 pulses with a power of 19 mW and pulse length of 600 ns were employed. The sample was placed on a Thorlabs piezo stage allowing for precise movements of the sample. The crystalline launching structures are created by placing multiple crystalline spots next to each other. The crystalline disks are created by placing multiple crystalline spots along a ring of the desired radius.

**s-SNOM**. In s-SNOM, a sharp metallic tip is brought into proximity to the sample surface. Upon illumination by a laser, the tip acts as an optical antenna leading to strongly enhanced near-fields at the tip's apex. The scattered light is detected interferometrically to extract amplitude and phase. We demodulate the signal at the second order of the tip-oscillation frequency. As s-SNOM is sensitive to local changes in the permittivity of the sample and to local near-fields[50], it can directly probe the near-fields of the polaritons[2–4,51,53]. We used a commercially available s-SNOM (NeaSNOM neaspec GmbH) to image the polaritons in $MoO_3$ on IST. We combined commercially available quantum cascade lasers (MIRcat-QT Mid-IR Laser by Daylight Solutions Inc.) with a liquid nitrogen-cooled Mercury Cadmium Tellurium detector to investigate polaritons in the frequency range from 900 to 945 $cm^{-1}$ and in the frequency range from 960 to 970 $cm^{-1}$ for the super-resolution regime. We used the pseudoheterodyne detection scheme to obtain s-SNOM amplitude and phase[49]. We performed our measurements using tapping amplitudes between 60 and 70 nm and demodulated the signal at the second demodulation order.

**Numerical Field Simulations.** Numerical simulations were performed with the commercial solver CST Studio Suite from Dassault Systèmes. The simulation was modelled with plane wave excitation using p-polarized light incident at angle $(\theta, \varphi) = (60°, 180°)$. Open boundaries with added free space padding where needed, were set.

The constant loss-free refractive index of 3.4 was used for Si. The dielectric function of IST is shown Figure 1b. A constant refractive index of 2.1 was assumed for the $ZnS:SiO_2$. The anisotropic dielectric function of $MoO_3$ was modelled according to Figure 1a.

The electric field was extracted 1 nm above the sample surface. The fields were normalized with respect to the incident field.

Effects of the SNOM tip were not considered in our simulations.

**Code Availability**

The computer codes developed for this study are available from the corresponding author upon request.

**Supporting Information**

Supplementary Notes (PDF) on a comparison of our fast fabrication of polariton optics to conventional fabrication schemes, on a comparative study of reconfigurable polariton optics and focusing structures, on in-plane hyperbolic polaritons, on polariton wavevector from s-SNOM line profiles, on super-resolution imaging, on tailoring confinement and field enhancement by disk distance variation on the enhancement and confinement along optical y-axis, on tailoring confinement and field enhancement by varying illumination frequency, and on the derivation of the dispersion relation of highly confined PhPs in the investigated layer stack.

**Acknowledgements**

The authors thank Maike Kreutz and Asli Turan for the sputter deposition of the thin film layer stack. L.J. acknowledges the support of RWTH University through the RWTH Graduate Support scholarship. L.J., L.C. and T.T. acknowledge support by the Deutsche Forschungsgemeinschaft (DFG No. 518913417 & SFB 917 "Nanoswitches"). M.W. acknowledges support by the Deutsche Forschungsgemeinschaft (SFB 917 "Nanoswitches"). E.T.-G. acknowledges support through the Severo Ochoa program from the Government of the Principality of Asturias (no. PA-23-PF-BP22-046). P.A.-G. acknowledges support from the European Research Council under Consolidator grant no. 101044461, TWISTOPTICS and the Spanish Ministry of Science and Innovation (State Plan for Scientific and Technical Research and Innovation grant number PID2022-141304NB-I00). G. Á.-P. acknowledges support from the European Union (Marie Skłodowska-Curie Actions, grant

**Author Contributions Statement**

L.J., L.C. and T.T. conceived the project with assistance by P.A.-G.. E.T.-G., C.L., and A.T.M.-L. contributed to preliminary sample fabrication, preliminary s-SNOM measurements. E.T.-G. performed the final sample fabrication. U.S. and A.M. performed the s-SNOM measurement. L.J. supervised the s-SNOM measurements. E.T.-G. assisted in the s-SNOM measurements and their interpretation. U.S. performed the numerical field simulations under supervision of L.C. L.J., U.S., A.M. and L.C. analyzed the data. G.A.-P. and C.L. provided theoretical assistance and support. P.A.-G. supervised E.T.-G., C.L., and A.T.M.-L.. M.W. provided sputtering equipment and phase-change material expertise. All authors contributed to writing the manuscript.

**Competing Interests Statement**

The authors declare no competing financial interest.

# Supplementary Information

# Fast Programming of In-Plane Hyperbolic Phonon Polariton Optics Through van der Waals Crystals using the Phase-Change Material In$_3$SbTe$_2$

Author(s) and Corresponding Author(s)*:

Lina Jäckering*[+,1,2], Umberto Saldarelli[+,1,2], Aaron Moos[1,2], Lukas Conrads[1,2], Enrique Terán-García[3,4], Christian Lanza[3,4], Aitana Tarazaga Martín-Luengo[3,4], Gonzalo Álvarez-Pérez[5], Pablo Alonso-González[3,4], Matthias Wuttig[1,2], Thomas Taubner*[,1,2]

jaeckering@physik.rwth-aachen.de

[+]both authors contributed equally

**Affiliations**

[1]1st Institute of Physics (IA), RWTH Aachen University, 52074 Aachen, Germany

[2]Juelich-Aachen Research Alliance (JARA-FIT), 52425 Juelich, Germany

[3]Department of Physics, University of Oviedo, 33006 Oviedo, Spain

[4]Center of Research on Nanomaterials and Nanotechnology, CINN (CSIC-Universidad de Oviedo), 33940 El Entrego, Spain

[5]Istituto Italiano di Tecnologia, Center for Biomolecular Nanotechnologies, Via Barsanti 14, 73010 Arnesano, Italy.

**This PDF file includes:**

**Supplementary Note 1: Comparison of our fast fabrication of polariton optics to conventional fabrication schemes**

**Supplementary Note 2: Comparative study of reconfigurable polariton optics and focusing structures**

**Supplementary Note 3: In-plane hyperbolic polaritons**

**Supplementary Note 4: Polariton wavevector from s-SNOM line profiles**

**Supplementary Note 5: Super-resolution imaging**

**Supplementary Note 6: Tailoring confinement and field enhancement by disk distance variation**

**Supplementary Note 7: Enhancement and confinement along optical y-axis**

**Supplementary Note 8: Tailoring confinement and field enhancement by varying illumination frequency**

**Supplementary Note 9: Derivation of the dispersion relation of highly confined PhPs in the investigated layer stack**

**Supplementary Note 1: Comparison of our fast fabrication of polariton optics to conventional fabrication schemes**

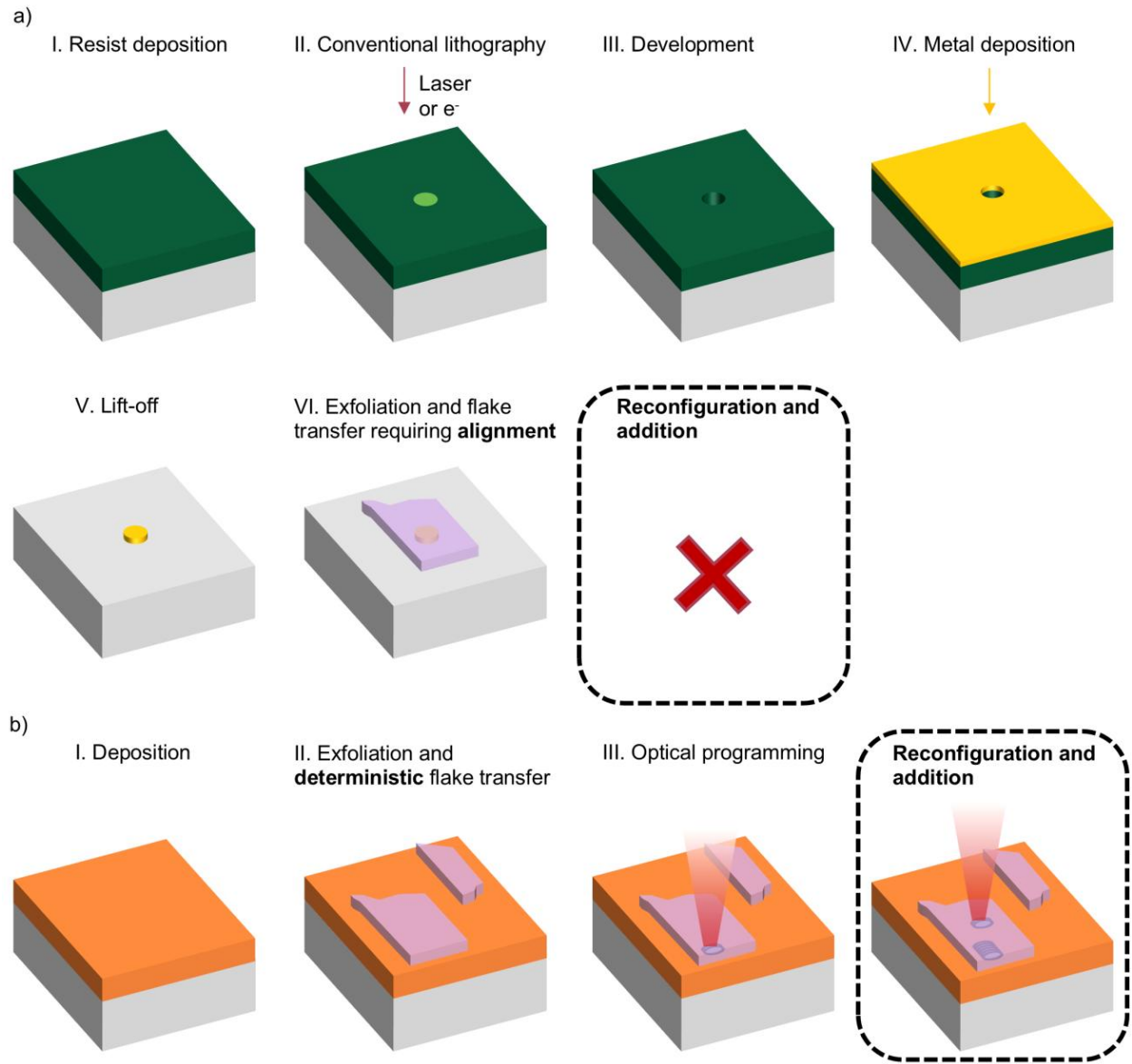

***Figure S1: Comparison of our fast fabrication of polariton optics to conventional fabrication schemes.*** *a) Conventional fabrication scheme of metallic polariton optics e.g. launching disks. Conventionally, metallic disks are fabricated via electron beam or optical lithography that involve the subsequent fabrication steps: (i) a resist mask is deposited on the substrate that is subsequently (II) patterned using an electron beam or a laser. Afterwards (III) the resist is developed and (IV) the metal is deposited. (V) In the lift-off the remaining resist with the deposited metal on top is removed leaving the aimed metal disk. To deposit an α-MoO₃ flake, first the flakes are exfoliated mechanically onto a different substrate. Then, the flake of choice is*

*transferred and aligned onto the previously fabricated metal disk. (VI) Afterwards, there is no option for adding another metal structure to the same flake nor an option for reconfiguration of the fabricated structure.  b) Our fast prototyping scheme bases on only three steps: First, (I) the amorphous IST film is deposited onto the substrate. Then, (II) α-MoO₃ flakes are directly exfoliated onto the amorphous IST. Finally, (III) a α-MoO₃ flake is chosen and the desired launching structures optically programmed through the α-MoO₃ flake. Thus, the desired structure can be aligned to the crystal axes flexibly. Our fabrication scheme allows us to add further launching structures to the same flake and also to reconfigure the previously fabricated structure by further crystallization.*

Conventionally, the fabrication of polariton launching structures relies on multi-process fabrication schemes such as electron beam or optical lithography[1]. To contrast our fast three step fabrication scheme with conventionally used, time-consuming lithography, we shortly outline the process steps for both fabrication schemes in Figure S1. In conventional lithography[1] (c.f. Figure S1a) the fabrication process starts with the deposition of a resist mask onto the substrate (I). Next, resist is patterned using an electron beam or laser (II). Afterwards, the resist is developed (III) and the metal deposited (IV). In a subsequent lift-off the remaining resist together with the metal on top of it is removed leaving the aimed metal launching structure (V). In order to align an α-MoO₃ flake with the fabricated metal structure, first, α-MoO₃ flakes are mechanically exfoliated onto a different substrate. Then, a α-MoO₃ flake is chosen, transferred and aligned with the metal structure (VI). Relying on these conventional techniques does not allow for adding another structure to the same flake nor for reconfiguration of the structure.

Our fabrication scheme is based on only three steps (c.f. Figure S1b). First, amorphous IST is deposited onto the substrate, followed by sputtering of a capping layer of $ZnS:SiO_2$ to prevent the IST from oxidation (omitted in sketches) (I). Secondly, α-MoO₃ flakes are mechanically exfoliated onto a polydimethylsiloxane (PDMS) stamp and deterministically transferred onto the thin IST film (III). Thirdly, a α-MoO₃ flake is chosen, and the desired launching structure is optically programmed through the α-MoO₃ by locally crystallizing the IST with a red laser (IV). Our fabrication scheme allows for adding another launching structure to the same flake and also for reconfiguration of the previously written structure.

To highlight the short turn-around time of our fabrication scheme compared to conventional fabrication schemes of lithography, in the following the overall time is estimated for both fabrication schemes: In conventional lithography, the deposition and patterning of the resist is estimated to several hours, followed by approximately 0.5 to 1 hours for the development, 1 to 2 hours for metal deposition and 0.5 hours for the lift off. Additionally, the exfoliation and subsequent transfer and alignment of an $\alpha$-MoO$_3$ flake to the structure takes several hours. These different process steps require different machines that are often located in different facilities. Taking into account the availabilities of the machines, often several days are required for fabrication.

In contrast, the overall time of our fabrication scheme can be estimated to be 1 to 2 hours for deposition of the IST film, 1 hour for mechanical exfoliation of the $\alpha$-MoO$_3$ flakes and 10 minutes for optical programming of the launching structures. Using our fabrication scheme, optical structures can be fabricated within a few hours. Moreover, additional structures for the same flake or different flakes on the same chip can be fabricated within another 10 minutes as the first two steps are not required. Hence, the use of fewer fabrication steps leads to faster turn-around times.

# Supplementary Note 2: Comparative study of reconfigurable polariton optics and focusing structures

| Publication | Pola-riton-hosting material | Tuning mechanism | Free propagation | Nanostructures | Focusing structure | |
|---|---|---|---|---|---|---|
| | | | | | tuning range focal point | Confine-ment |
| This work | $\alpha$-MoO$_3$ | Phase change in IST<br>- Non-volatile<br>- Reconfigurable<br>- Modification of size and shape | **Launching at phase boundary** | **Launching focusing** structure<br><br>**nanoresonator** | 0.55-2.14 µm (frequency variation) | x: $\lambda_0$/28; $\lambda_p$/2.1<br>y: $\lambda_0$/10; $\lambda_p$/0.8<br><br>x: $\lambda_0$/35; $\lambda_p$/2.7<br>y: $\lambda_0$/20; $\lambda_p$/1.37 |
| Jäckering et al.[2] | hBN | Phase change in IST<br>- Non-volatile<br>- Reconfigurable<br>- Modification of size and shape | **Launching at phase boundary** | Circular **resonator**<br><br>**Launching focusing** structure | 1.3-3.4 µm (curvature tuning) | |
| Lu et al. [3] | h$^{10}$BN<br><br>$\alpha$-MoO$_3$ | Phase change in IST<br>- Non-volatile<br>- Reconfigurable | Free propagation of HPhP in hBN and $\alpha$-MoO3 tailored by IST phases | **launching and focussing** antenna for HPhP in $\alpha$-MoO3 | ~1-2.8 µm (extracted from Figure 4F) (curvature + frequency tuning) | - |
| Aghamiri et al. [4] | h$^{10}$BN<br><br>$\alpha$-MoO$_3$ | Tune charge carrier concentration in SmNiO$_3$ (SNO) via<br>- strong local fields: non-volatile<br>- hydrogenation: non-volatile, global<br>- temperature: volatile, global | Free propagation of HPhP in hBN and $\alpha$-MoO$_3$ tailored by tuned local conductivity in SNO; | Triangular **cavity** for HPhP in hBN (not reconfigure nor tailored)<br><br>Circular, rectangular **resonator** for HPhP in $\alpha$-MoO$_3$ (not reconfigured nor tailored) | - | - |
| Qu et al.[5] | $\alpha$-MoO$_3$ | $\alpha$-MoO$_3$ on graphene: Tuning of fermi energy in graphene volatile | | **Launching focusing** structure | ~0.7-7.4 µm) (tuning by disk diameter variation, frequency tuning)<br>~2.8-4.6 µm (extracted from Figure 4f) (by tuning fermi energy in graphene | x: -<br>y: $\lambda_0$/33 |
| Hu et al. [6] | $\alpha$-MoO$_3$ | $\alpha$-MoO$_3$ on graphene: Tuning of | | **Launching focusing** structure | - | x: -<br>y: $\lambda_0$/21; |
| | | fermi energy in graphene volatile | | | | |
| Liang et al. [7] | α-MoO$_3$ | α-MoO$_3$ Te-thin film: variation of Te-film thickness | | **Launching focusing** structure | 2.29-5.61 µm | x: - y: $\lambda_0$/20; $\lambda_p$/1.8 ($\lambda_p$ extracted from Figure 4b) |
| Martín-Sánchez et al. [8] | α-MoO$_3$ | No tuning | | **Launching focusing** structure Improved shape | 0.6-1.7 µm (tuning frequency) | x: $\lambda_0$/28 y: $\lambda_0$/34; $\lambda_p$/2 x: y: $\lambda_0$/50; $\lambda_p$/4.5 |
| Zheng et al. [9] | α-MoO$_3$ | No tuning | | **Launching focusing** structure | 0.6-1.6 µm (extracted from Figure 4G) | x: $\lambda_0$/32 y: $\lambda_0$/30; $\lambda_p$/1.04 ($\lambda_p$ extracted from Figure 4G) |

**Table S1: Comparative study of reconfigurable polariton optics and focusing structures**

A comparative study of reconfigurable polariton optics for polaritons in α-MoO$_3$ and polariton focusing structures in related literature is summarized in Table S1. We highlight the reconfigurability and compare the tuning range and achieved confinement of focusing structures.

**Comparison with previously investigated reconfigurable or tunable polariton optics**

Tunable and reconfigurable optics for polaritons in α-MoO$_3$ have been previously achieved by substrate-engineering, i.e. the substrate of the α-MoO$_3$ is changed and thereby the polariton propagation is altered. Liang et al.[7] combined α-MoO$_3$ with thin-films of Te of different thicknesses. Relying on conventional lithography techniques, a focusing gold disk was added, and by fabricating samples of different Te-film thicknesses the focal length was tuned. However, this approach still requires conventional fabrication techniques and film thicknesses cannot be changed afterwards, thus this approach is not suitable for fast prototyping nor reconfiguring of polariton optics. Tunable focusing structures for polariton in α-MoO$_3$ have been realized by combing α-MoO$_3$ with graphene.[5,6] Graphene's optical properties are gate tunable and therefore also the polariton propagation in the α-MoO$_3$ above can be tailored upon gating. Thereby, Hu et al.[6] and Qu et al.[5] realized gate-tunable focusing with conventionally fabricated gold-launching structures. Fast prototyping of polariton optics such as circular and rectangular resonators before α-MoO$_3$ flake deposition has been achieved by Aghamiri et al.[4] by combing α-MoO$_3$ with the oxide SNO. The conductivity and thus also the optical properties of SNO can be locally tuned by applying high electric fields and thus polariton optics can directly be imprinted. Recently, Lu et al.[3] combined α-MoO$_3$ with the PCM IST and demonstrated tailoring of the polariton propagation by the two IST phases. Further, they transferred a α-MoO$_3$ flake onto crystalline antennas of different widths programmed into the IST before. These crystalline antennas focus the polaritons. However, programming of polariton optics after flake deposition and their reconfiguration have not been demonstrated using SNO nor IST. In recent work by our group [2], fast prototyping and reconfiguring of a focusing arc structure programmed for in-plane isotropic phonon polaritons in hBN into IST has been demonstrated. In this work, we follow-up on the previous approach and combine α-MoO$_3$ with IST to tailor the in-plane hyperbolic phonon polaritons in α-MoO$_3$. We optically program launching structures through α-MoO$_3$ into the IST. As we can optically program the structures after flake deposition, we demonstrate a fast prototyping scheme. Furthermore, we show reconfiguration of already fabricated structures.

**Comparison with previously investigated static polariton focusing structures**

In our work, we tune the focal length of a single focusing structure from 0.55 to 2.14 μm by variation of the illumination frequency and show a confinement of $\lambda_0/28 = \lambda_p/2.1$ along the optical x-axis and of $\lambda_0/10 = \lambda_p/0.8$ along the optical y-axis. By adding a second focusing disk we increase the confinement to $\lambda_0/35 = \lambda_p/2.7$ along the optical x-axis and to $\lambda_0/20 = \lambda_p/1.37$ along the optical y-axis. Our achieved tuning range of the focal length (0.55 to 2.14 μm) is slightly larger than that achieved by Martín-Sánchez et al. [8] (0.6-1.7 μm) and Zheng et al.[9] (0.6-1.6 μm) who relied on conventionally fabricated focusing structures. The conventionally fabricated focusing structures of different radii by Qu et al.[5] allowed for a slightly larger tuning range of 0.7 to 7.4 µm. The confinement of $\lambda_0/28 = \lambda_p/2.1$ along the optical x-axis and of $\lambda_0/10 = \lambda_p/0.8$ along the optical y-axis achieved with our focusing structures is comparable to that observed for conventionally fabricated focusing structures by Zheng et al.[9] ($\lambda_0/32$ along the optical x-axis and $\lambda_0/30 = \lambda_p/1.04$ along the optical y-axis) and slightly lower than that by Qu et al.[5] ($\lambda_0/33$ along the optical y-axis) and by Martín-Sánchez et al.[8] ($\lambda_0/28$ along the optical x-axis and $\lambda_0/34 = \lambda_p/2$ along the optical y-axis). However, the confinement achieved with the double disk structure ($\lambda_0/35 = \lambda_p/2.7$ along the optical x-axis and $\lambda_0/20 = \lambda_p/1.37$ along the optical y-axis) compares well with the higher values observed in literature. Only the focusing structure optimized to increase the contribution of high wavevectors and decrease that of low wave vector by Martín-Sánchez et al.[8] shows a higher confinement ($\lambda_0/50 = \lambda_p/4.5$ along the optical y-axis). In summary, our work presents fast-prototyping of reconfigurable focusing structures for polaritons in $\alpha$-MoO$_3$ that do show a performance in terms of tuning range of the focal length and confinement comparable to previously investigated, conventionally fabricated focusing structures.

**Supplementary Note 3: In-plane Hyperbolic Polaritons**

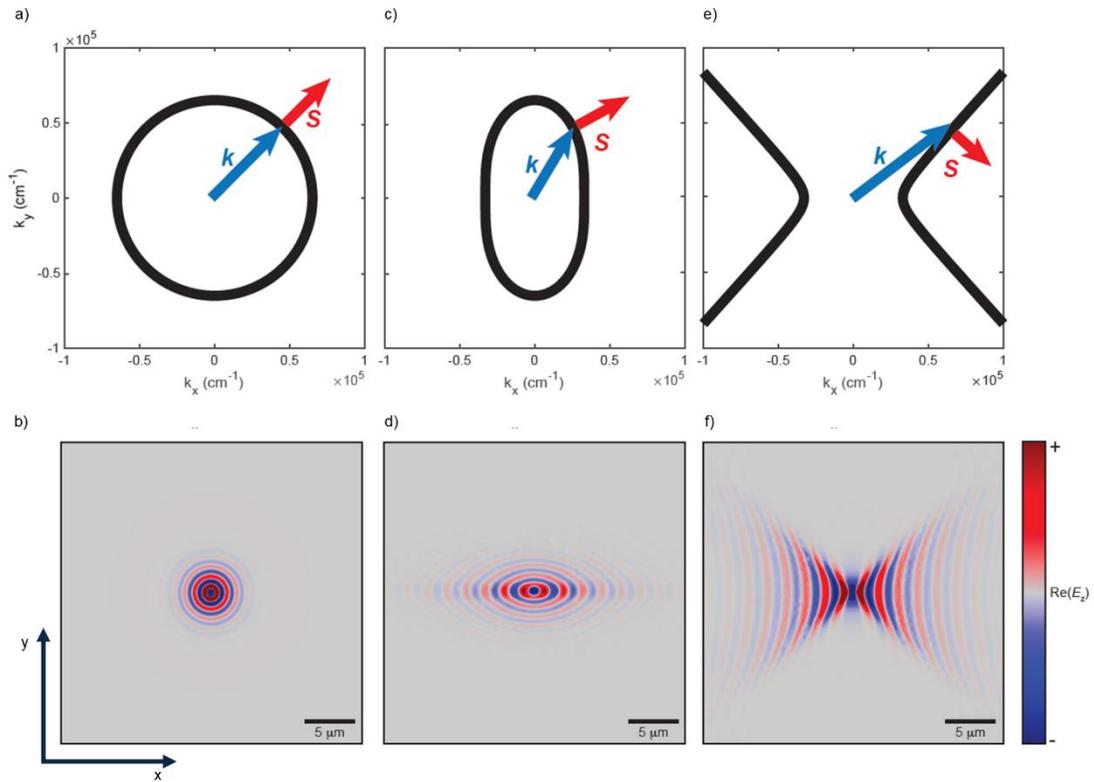

***Figure S2: In-plane Hyperbolic Polaritons.*** *a) IFC in Fourier space ($k_x,k_y$) at fixed $k_z$ and b) sketch of the in-plane propagation of in-plane isotropic polaritons. c) IFC in Fourier space ($k_x,k_y$) and d) sketch of the in-plane propagation of in-plane elliptical polaritons. e) IFC in Fourier space ($k_x,k_y$) and f) sketch of the in-plane propagation of in-plane elliptical polaritons. The red-dashed line and the red solid line indicate the polariton wave vector and the Poynting vector, respectively.*

In the case of in-plane isotropic media, the permittivity tensor components along the optical x- and y-axes have the same value. Thus, the isofrequency contour (IFC) which can directly be retrieved from the dispersion equation resembles a circle in the Fourier space ($k_x,k_y$) at fixed $k_z$ as sketched in Figure S2a.[10,11] As the Poynting vector is parallel to the normal of the IFC, both the Poynting vector and the wave vector are collinear for in-plane isotropic materials as can be observed in Figure S2a.[10,11] The circular IFC leads to an in-plane isotropic propagation of polaritons as sketched in Figure S2b.[10,11]

In the case of in-plane elliptical media, the permittivity along the optical x- and y-axes is different but of the same sign. Thus, the IFC resembles an ellipse (c.f. Figure S2c) and the Poynting vector and the wave vector are not collinear in general.[11] The elliptical IFC leads to an in-plane anisotropic propagation in real space (c.f. Figures S2d).[11]

In the case of in-plane hyperbolic media, the permittivity along the optical x- and y-axes is different and also opposite in sign. This leads to an in-plane hyperbolic IFC.[10–12] Figure S2e sketches the hyperbolic IFC in Fourier space ($k_x$,$k_y$) for the case of a negative permittivity along the optical x-axis. From the sketch, we can see that both the Poynting vector and the wave vector are not collinear.[10,11] Furthermore, the IFC restricts the direction of wave vectors. Only wave vectors with an angle γ with respect to the optical x-axis smaller than the IFC opening angle α are allowed.[10,11] This leads to the highly directional propagation polaritons in in-plane hyperbolic media as sketched in Figure S2f.[10,13]

**Supplementary Note 4: Polariton Wavevector from s-SNOM Line Profiles**

For a quantification of the polariton wavevector in Figure 1 and Figure 2 in the main text, we fitted line profiles extracted from s-SNOM amplitude images as described in the following. In s-SNOM there are typically two launching mechanisms of freely propagating polaritons: tip-launching and edge-launching. The tip-launched polaritons are launched by the sharp s-SNOM tip and reflected, e.g. at a sample edge[14–16], leading to an interference pattern with a fringe spacing of $\delta = \frac{\lambda_p}{2} = \frac{\Pi}{k_p}$.[14–17] The edge-launched polaritons are launched at a discontinuity, such as an edge, a phase-boundary or a defect and interfere with the incident light[17,18]. The resulting interference pattern shows a fringe spacing of $\delta \approx \lambda_p = \frac{2\Pi}{k_p}$ when $k_p$>>$k_0$[17,18]. In our experiments, we do only observe contributions of polaritons launched at the phase-boundary of IST (edge-launched). We use the following fit function for edge-launched polaritons, considering the geometric decay of $1/x^a$[17], where a has been found to be a~1 in previous studies[17], and a linear background in Figure 1, (in Figure 2: C=0):

$$s \sim |A\frac{e^{ik_p(x-x_0)}}{x - x_0} + B + Cx|$$

Here, A, B, C, and $x_0$ are free parameters.

**Supplementary Note 5: Super-resolution imaging**

As we program the cIST structures after having exfoliated $\alpha$-MoO$_3$ we cannot characterize its exact size and shape by direct measures (e.g. by atomic force microscopy as it is done for conventionally fabricated structures). We overcome this issue by performing a super-resolution s-SNOM amplitude image by tuning our laser source to 970 cm$^{-1}$, close to the lower bound of the upper reststrahlenband $\alpha$-MoO$_3$. The super-resolution image of the cIST disk is shown in Figure 3b. We observe a circular area of high amplitude (surrounded by a white dashed contour-line) in the center of the image. Outside the circular area the amplitude is approximately constant and lower. In the upper reststrahlenband both in-plane permittivities are positive ($\epsilon_x$>0, $\epsilon_y$>0) whereas the out-of-plane permittivity is negative ($\epsilon_z$<0).[13] Thus, the propagation behavior is similar to that observed for HPhP in hBN.[19,20] The hyperbolic PhP in $\alpha$-MoO$_3$ propagate under a restricted angle through the volume of the $\alpha$-MoO$_3$ slab. At the lower bound of the upper reststrahlenband, the PhP in $\alpha$-MoO$_3$ will propagate under an angle close to zero, allowing for a direct image of the structure below, the so-called super-resolution imaging. The super-resolution imaging is similar to that observed in hBN, where super-resolution imaging was exploited to visualize gold-disks[19–21] and defects in buried few-layer graphene flakes[22,23]. Previously, canalization based super-resolution imaging of buried gold disks with a $\alpha$-MoO$_3$ SiC heterostructure has been demonstrated[24]. Therefore, we interpret the circular structure of high amplitude as a direct image of the cIST disk below the $\alpha$-MoO$_3$ flake. In the middle of the bright circle in Figure 3b, we observe a feature having a lower amplitude. The feature of low amplitude in the center aligns with the dark spot visible in the center of the bright crystalline area in the light microscope image (Figure 3a). Most likely, this feature corresponds to a slightly burnt spot during optical programming. We use this direct image to extract size, shape and position of the cIST disk. The crystallized circle has a slightly elongated axis along the y-direction.

**Supplementary Note 6: Enhancement and confinement along optical y-axis**

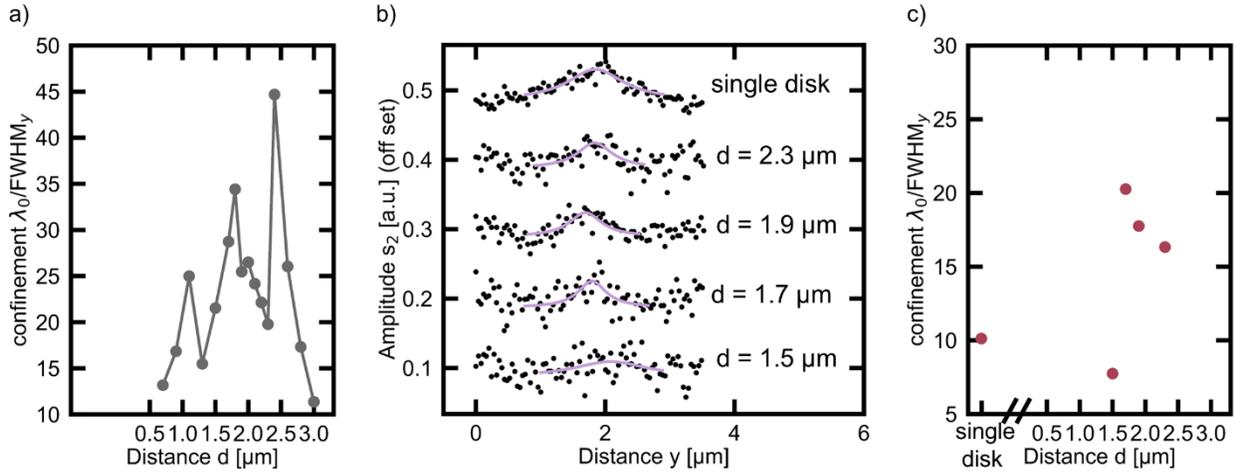

***Figure S3: Enhancement and confinement along optical y-axis.*** *a) Field confinement $\lambda_0/FWHM_y$ along the optical y-axis as function of disk distance extracted from simulations such as displayed in Figure 4d) in the main text. b) Line profiles of a single launching disk (top) extracted from s-SNOM images in Figure 3d) in main text and line profiles of two launching disks whose distance increases from top to bottom extracted from s-SNOM images in Figure 4f) in main text. c) Confinement $\lambda_0/FWHM_y$ along y-direction determined from line profiles extracted from experiment (red dots) as a function of disk distance.*

To quantify the influence of the disk variation on the confinement along the y-direction analogous to the x-direction discussed in Figure 4, we extract line profiles from the simulations (c.f. green dashed line in sketch in Figure 4b in main text) and determine the $FWHM_y$ of the maximum closest to the middle of the two disks to calculate the confinement $\lambda_0/FWHM_y$. Confinement along the y-direction is displayed as function of the disk distance extracted from simulations in Figure S3a. The maximal confinement along the y-direction corresponds to $\lambda_0/45$ and is found at a disk distance of 2.4 µm.

To experimentally quantify the confinement along the y-direction, we extract line profiles along y-direction from the s-SNOM amplitude images displayed in Figure 4f through the maximum closest to the middle of the two disks (c.f. green dashed line in Figure 4b in main text; for the single disk closest to the focal point) and compare them to a line profile extracted for a single launching disk in Figure S3b. We determine the $FWHM_y$ of the maximum by fitting a Lorentzian function. The experimental confinement along the y-direction is plotted in Figure S3c. Along the y-direction we find an enhanced confinement of $\lambda_0/20 = \lambda_p/1.4$ at a disk distance of 1.7 μm. Compared to the confinement of the single disk along the same axis ($\lambda_0/10 = \lambda_p/0.8$) the double disk arrangement allows for an increase in the confinement by factor of 2.

**Supplementary Note 7: Tailoring Confinement and Field Enhancement by disk distance variation**

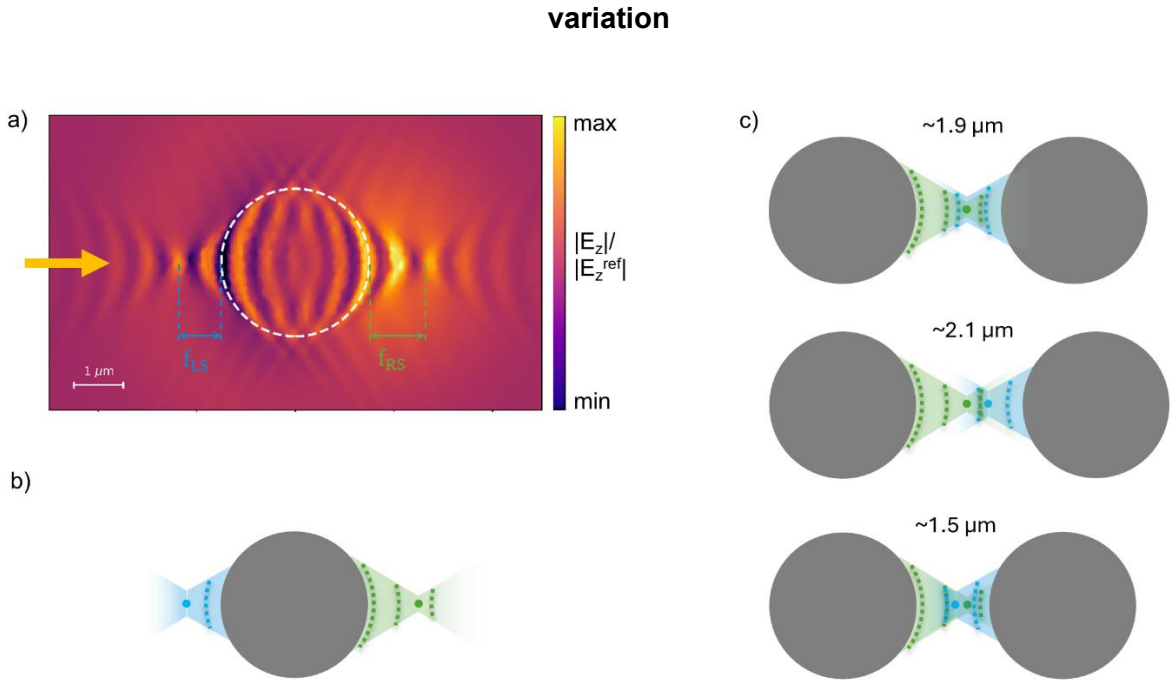

***Figure S4: Interference of polaritons launched by two opposing focusing disks.*** *a) Simulation of a single launching disk at 930 cm⁻¹. The illumination is considered to be from the left side. The scale bar is 1 µm. b) Sketch of polaritons launched and focused to both sides of the crystalline disk. c) Sketch of the interference of polaritons launched by two opposing focusing disks for different disk distances. The wave fronts are indicated by dashed curved lines and the focusing is indicated by the shaded area. The focal points are indicated by a dot. Focusing towards the left is sketched in blue and to the right in green.*

To understand the interference of polaritons launched by two opposing focusing disks, we first revisit the simulation of polaritons launched by a single disk as depicted in Figure S4a. Polaritons are launched towards the left and right. On both sides the polaritons are focused on a focal point. However, due to the considered illumination direction from the left (c.f. yellow arrow in Figure S4a), which was chosen to resemble the experiments, the focusing is not symmetric. We observe a shorter focal length on the left side of the disk than on the right side due to the interference with the incident light. Furthermore, the polaritons are launched with opposing phases as on the left side there is a minimum at the disk boundary whereas on the right side there is a maximum at the disk boundary. For simplification, we sketch the different behaviors of polaritons launched to left (green) and right (blue) in Figure S4b. The envelope of the focusing is indicated by a shaded cone and the wave fronts of the polaritons are indicated by dashed lines. The focal points are indicated by dots.

If we now consider the interference of polaritons launched by two opposing disks, polaritons launched to the right interfere with polaritons launched to the left by the other disk (c.f. sketches in Figure S4c). Although both disks have the same radius, their focal length are different due to the illumination direction. The asymmetry observed for the single disk determines the interference of the polaritons and thus the distances of maximal field confinement and enhancement. We find a maximal field confinement at a disk distance of approximately 1.9 μm (see top sketch in Figure S4c). At this disk distance the two focal points overlap. However, due to the opposing launching phases, there is not a constructive interference at the focal point similar to the observations in ref.[11] for an hyperbolic nanocavity. Hence, the field enhancement is not maximal. Instead, maximal field enhancement is observed at slightly larger or lower distances where constructive interference is achieved, however the focal points do not overlap (c.f. sketches in the bottom of Figure S4c).

**Supplementary Note 8: Tailoring Confinement and Field Enhancement by varying illumination frequency**

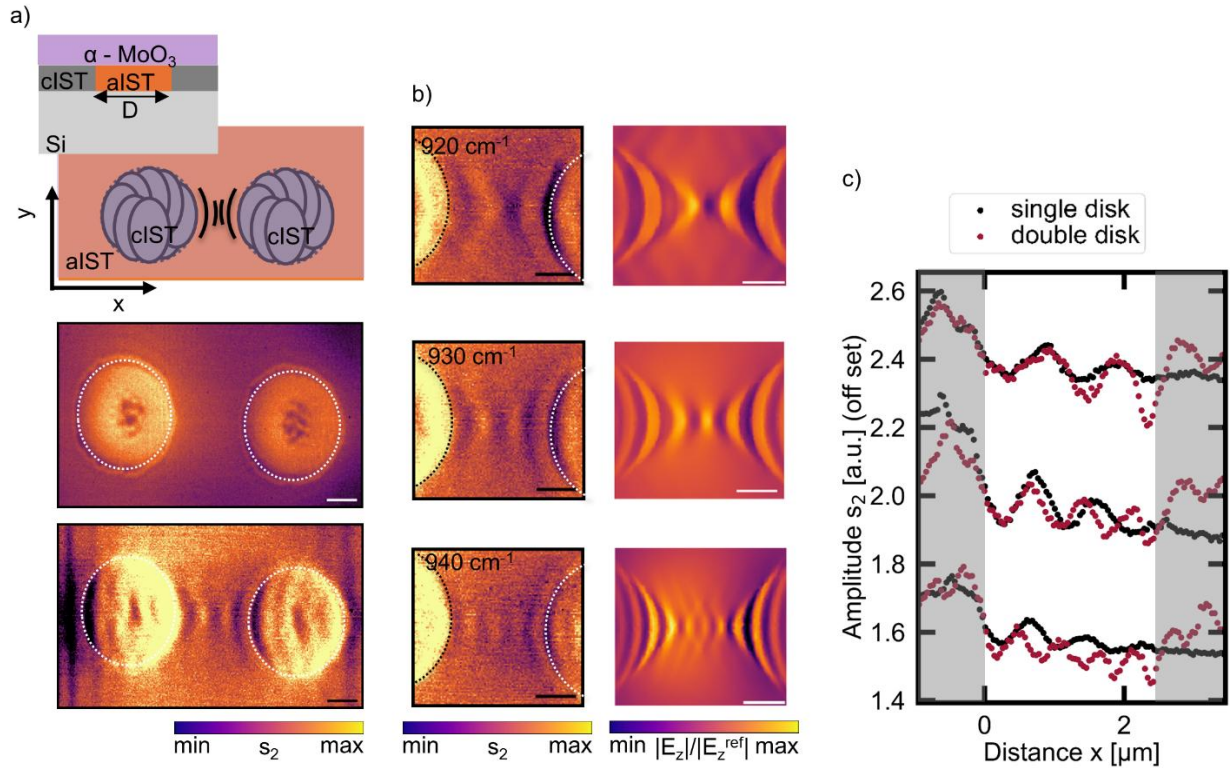

***Figure S5: Optically programmed focusing structures with matching focal point to confine phonon polaritons in α-MoO₃. a)*** *Sketch of the two circular crystalline launching structures of distance 1.9 µm programmed into IST below α-MoO₃ as side (top) and top (middle) view together with an s-SNOM amplitude image at 970 cm⁻¹(top) and at 930 cm⁻¹(bottom). b) s-SNOM amplitude images and simulations of the two focusing disks for three different frequencies increasing from top to bottom. The scale bars correspond to 1 µm. c) Line profiles of a single launching disk (in black) extracted from s-SNOM images in Figure 2d) and line profiles of two launching disks whose focal points overlap (in red) extracted from s-SNOM images in b).*

Tailoring of confinement and enhancement of polaritons between two disks can also be achieved by tuning the incident illumination frequency. We imaged the double disk structure with a distance of 1.9 µm shown in Figure 4 in the main text at three different frequencies and compare it to the respective simulations (c.f. Figure S5b). As the polariton wavelength and the focal length depend on the frequency also the interference pattern is tailored by changing the frequency. For quantification we extract line profiles and compare them to line profiles extracted from single disk at the same frequencies in Figure S5c. With increasing frequency, the polariton wavelength decreases. Consequently, in the case of the double disks we observe an increasing number of maxima between the disks. At 920 $cm^{-1}$ two maxima can be observed, at 930 $cm^{-1}$ three and at 940 $cm^{-1}$ four maxima. Comparing single and double disk for all frequencies the maxima of the single disks appear broader than that of the double disks indicating an increased confinement in case of the double disks.

## Supplementary Note 9: Derivation of the dispersion relation of highly confined PhPs in the investigated layer stack

In this section we show how to obtain the dispersion relation of highly confined polaritons in the used layer stack comprising of the silicon substrate (0), the amorphous IST thin film (1), the ZnS:SiO$_2$ capping layer (2), the α-MoO$_3$ flake (3), and air as superstrate (4). Since in the high-momentum approximation the contribution of the ordinary modes can be neglected [25], we only consider $\boldsymbol{p}$ modes for the superstrate and substrate and $\boldsymbol{e}$ modes for the biaxial layers. Following the same notation as in ref.[25], the boundary conditions for the tangential components of the electric and magnetic fields can be written as follows:

$$a_p^{4\uparrow}e^{iq_{4,p,z}(d_1+d_2+d_3)}\boldsymbol{p}_4 = \left(a_{e_2}^{3\uparrow}e^{iq_{3,e,z}(d_1+d_2+d_3)} + a_{e_2}^{3\downarrow}e^{-iq_{3,e,z}(d_1+d_2+d_3)}\right)\boldsymbol{e}_3$$
$$a_p^{4\uparrow}e^{iq_{4,p,z}(d_1+d_2+d_3)}\Upsilon_p^4\boldsymbol{p}_{4,+} = a_{e_2}^{3\uparrow}e^{iq_{3,e,z}(d_1+d_2+d_3)}\boldsymbol{e'}_{3,+} + a_{e_2}^{3\downarrow}e^{-iq_{3,e,z}(d_1+d_2+d_3)}\boldsymbol{e'}_{3,-} \tag{S2}$$

$$\left(a_{e_2}^{3\uparrow}e^{iq_{3,e,z}(d_1+d_2)} + a_{e_2}^{3\downarrow}e^{-iq_{3,e,z}(d_1+d_2)}\right)\boldsymbol{e}_3 = \left(a_{e_2}^{2\uparrow}e^{iq_{2,e,z}(d_1+d_2)} + a_{e_2}^{2\downarrow}e^{-iq_{2,e,z}(d_1+d_2)}\right)\boldsymbol{e}_2$$
$$a_{e_2}^{3\uparrow}e^{iq_{3,e,z}(d_1+d_2)}\boldsymbol{e'}_{3,+} + a_{e_2}^{3\downarrow}e^{-iq_{3,e,z}(d_1+d_2)}\boldsymbol{e'}_{3,-} = a_{e_2}^{2\uparrow}e^{iq_{2,e,z}(d_1+d_2)}\boldsymbol{e'}_{2,+} + a_{e_2}^{2\downarrow}e^{-iq_{2,e,z}(d_1+d_2)}\boldsymbol{e'}_{2,-} \tag{S3}$$

$$\left(a_{e_2}^{2\uparrow}e^{iq_{2,e,z}d_1} + a_{e_2}^{2\downarrow}e^{-iq_{2,e,z}d_1}\right)\boldsymbol{e}_2 = \left(a_{e_2}^{1\uparrow}e^{iq_{1,e,z}d_1} + a_{e_2}^{1\downarrow}e^{-iq_{1,e,z}d_1}\right)\boldsymbol{e}_1$$
$$a_{e_2}^{2\uparrow}e^{iq_{2,e,z}d_1}\boldsymbol{e'}_{2,+} + a_{e_2}^{2\downarrow}e^{-iq_{2,e,z}d_1}\boldsymbol{e'}_{2,-} = a_{e_2}^{1\uparrow}e^{iq_{1,e,z}d_1}\boldsymbol{e'}_{1,+} + a_{e_2}^{1\downarrow}e^{-iq_{1,e,z}d_1}\boldsymbol{e'}_{1,-} \tag{S4}$$

$$\left(a_{e_2}^{1\uparrow} + a_{e_2}^{1\downarrow}\right)\boldsymbol{e}_1 = a_p^{0\downarrow}\boldsymbol{p}_0$$
$$a_{e_2}^{1\uparrow}\boldsymbol{e'}_{1,+} + a_{e_2}^{1\downarrow}\boldsymbol{e'}_{1,-} = a_p^{0\downarrow}\Upsilon_\beta^0\boldsymbol{p}_0 \tag{S5}$$

Eqs. (S2-S5) stand for the boundary conditions at the interfaces between the superstrate (4) and the top (3) layer, between the top (3) and the middle (2) layers, between the middle (2) and the bottom (1) layers, and between the bottom (1) layer and the substrate (0), respectively. Next, we project Eqs. (S2-S5) onto the subvectors $\boldsymbol{p}$. Considering the approximations $\boldsymbol{p} \cdot \boldsymbol{e}_j = 1$ and $\boldsymbol{p} \cdot \boldsymbol{e'}_{j,\pm} = \frac{\varepsilon_{j,z}q_{j,e,z}}{\pm iq_{j,o,z}^2}$, we rewrite Eqs. (S2-S5) as follows:

$$a_p^{4\uparrow}e^{iq_{4,p,z}(d_1+d_2+d_3)} = a_{e_2}^{3\uparrow}e^{iq_{3,e,z}(d_1+d_2+d_3)} + a_{e_2}^{3\downarrow}e^{-iq_{3,e,z}(d_1+d_2+d_3)}$$
$$-a_p^{4\uparrow}e^{iq_{4,p,z}(d_1+d_2+d_3)}\frac{\varepsilon_4}{iq} = \frac{\varepsilon_{3,z}q_{3,e,z}}{iq^2}\left(a_{e_2}^{3\uparrow}e^{iq_{3,e,z}(d_1+d_2+d_3)} - a_{e_2}^{3\downarrow}e^{-iq_{3,e,z}(d_1+d_2+d_3)}\right) \tag{S6}$$

$$a_{e_2}^{3\uparrow}e^{iq_{3,e,x}(d_1+d_2)} + a_{e_2}^{3\downarrow}e^{-iq_{3,e,x}(d_1+d_2)} = a_{e_2}^{2\uparrow}e^{iq_{2,e,x}(d_1+d_2)} + a_{e_2}^{2\downarrow}e^{-iq_{2,e,x}(d_1+d_2)}$$

$$\frac{\varepsilon_{3,z}q_{3,e,z}}{iq^2}\left(a_{e_2}^{3\uparrow}e^{iq_{3,e,x}(d_1+d_2)} - a_{e_2}^{3\downarrow}e^{-iq_{3,e,x}(d_1+d_2)}\right) = \frac{\varepsilon_{2,z}q_{2,e,z}}{iq^2}\left(a_{e_2}^{2\uparrow}e^{iq_{2,e,x}(d_1+d_2)} - a_{e_2}^{2\downarrow}e^{-iq_{2,e,x}(d_1+d_2)}\right) \quad \text{(S7)}$$

$$a_{e_2}^{2\uparrow}e^{iq_{2,e,x}d_1} + a_{e_2}^{2\downarrow}e^{-iq_{2,e,x}d_1} = a_{e_2}^{1\uparrow}e^{iq_{1,e,x}d_1} + a_{e_2}^{1\downarrow}e^{-iq_{1,e,x}d_1}$$

$$\frac{\varepsilon_{2,z}q_{2,e,z}}{iq^2}\left(a_{e_2}^{2\uparrow}e^{iq_{2,e,x}d_1} - a_{e_2}^{2\downarrow}e^{-iq_{2,e,x}d_1}\right) = \frac{\varepsilon_{1,z}q_{1,e,z}}{iq^2}\left(a_{e_2}^{1\uparrow}e^{iq_{1,e,x}d_1} - a_{e_2}^{1\downarrow}e^{-iq_{1,e,x}d_1}\right) \quad \text{(S8)}$$

$$a_{e_2}^{1\uparrow} + a_{e_2}^{1\downarrow} = a_p^{0\downarrow}$$

$$\frac{\varepsilon_{1,z}q_{1,e,z}}{iq^2}\left(a_{e_2}^{1\uparrow} + a_{e_2}^{1\downarrow}\right) = a_p^{0\downarrow}\frac{\varepsilon_0}{iq} \quad \text{(S9)}$$

For convenience, we multiply the second boundary condition in Eqs. (S6-S9) by $iq^2$, leading to the following system of equations:

$$a_p^{4\uparrow}e^{iq_{4,p,x}(d_1+d_2+d_3)} = a_{e_2}^{3\uparrow}e^{iq_{3,e,x}(d_1+d_2+d_3)} + a_{e_2}^{3\downarrow}e^{-iq_{3,e,x}(d_1+d_2+d_3)}$$

$$-a_p^{4\uparrow}e^{iq_{4,p,x}(d_1+d_2+d_3)}\varepsilon_4 q = \varepsilon_{3,z}q_{3,e,z}\left(a_{e_2}^{3\uparrow}e^{iq_{3,e,x}(d_1+d_2+d_3)} - a_{e_2}^{3\downarrow}e^{-iq_{3,e,x}(d_1+d_2+d_3)}\right) \quad \text{(S10)}$$

$$a_{e_2}^{3\uparrow}e^{iq_{3,e,x}(d_1+d_2)} + a_{e_2}^{3\downarrow}e^{-iq_{3,e,x}(d_1+d_2)} = a_{e_2}^{2\uparrow}e^{iq_{2,e,x}(d_1+d_2)} + a_{e_2}^{2\downarrow}e^{-iq_{2,e,x}(d_1+d_2)}$$

$$\varepsilon_{3,z}q_{3,e,z}\left(a_{e_2}^{3\uparrow}e^{iq_{3,e,x}(d_1+d_2)} - a_{e_2}^{3\downarrow}e^{-iq_{3,e,x}(d_1+d_2)}\right) = \varepsilon_{2,z}q_{2,e,z}\left(a_{e_2}^{2\uparrow}e^{iq_{2,e,x}(d_1+d_2)} - a_{e_2}^{2\downarrow}e^{-iq_{2,e,x}(d_1+d_2)}\right) \quad \text{(S11)}$$

$$a_{e_2}^{2\uparrow}e^{iq_{2,e,x}d_1} + a_{e_2}^{2\downarrow}e^{-iq_{2,e,x}d_1} = a_{e_2}^{1\uparrow}e^{iq_{1,e,x}d_1} + a_{e_2}^{1\downarrow}e^{-iq_{1,e,x}d_1}$$

$$\varepsilon_{2,z}q_{2,e,z}\left(a_{e_2}^{2\uparrow}e^{iq_{2,e,x}d_1} - a_{e_2}^{2\downarrow}e^{-iq_{2,e,x}d_1}\right) = \varepsilon_{1,z}q_{1,e,z}\left(a_{e_2}^{1\uparrow}e^{iq_{1,e,x}d_1} - a_{e_2}^{1\downarrow}e^{-iq_{1,e,x}d_1}\right) \quad \text{(S12)}$$

$$a_{e_2}^{1\uparrow} + a_{e_2}^{1\downarrow} = a_p^{0\downarrow}$$

$$\varepsilon_{1,z}q_{1,e,z}\left(a_{e_2}^{1\uparrow} + a_{e_2}^{1\downarrow}\right) = a_p^{0\downarrow}\varepsilon_0 q \quad \text{(S13)}$$

Next, we start from Eq. (S13) to iteratively solve the system. By dividing the top and bottom boundary conditions in Eq. (S13), we get to

$$a_{\gamma_1}^{1\downarrow} = \left(\frac{\varepsilon_{1,z}q_{1,e,z} - \varepsilon_s q}{\varepsilon_{1,z}q_{1,e,z} + \varepsilon_s q}\right)a_{\gamma_1}^{1\uparrow} \quad \text{(S14)}$$

Inserting Eq. (S14) into Eq. (S12), and dividing the two boundary conditions, we get to

$$\frac{a_{e_2}^{2\uparrow}e^{iq_{2,e,x}d_1} + a_{e_2}^{2\downarrow}e^{-iq_{2,e,x}d_1}}{\varepsilon_{2,z}q_{2,e,z}\left(a_{e_2}^{2\uparrow}e^{iq_{2,e,x}d_1} - a_{e_2}^{2\downarrow}e^{-iq_{2,e,x}d_1}\right)} = D_1 \quad \text{(S15)}$$

where

$$D_1 = \frac{\varepsilon_{1,z}q_{1,e,z}cos(\xi_{1,e,z}) + i\varepsilon_0 q sin(\xi_{1,e,z})}{\varepsilon_{1,z}q_{1,e,z}\varepsilon_0 q cos(\xi_{1,e,z}) + i(\varepsilon_{1,z}q_{1,e,z})^2 sin(\xi_{1,e,z})} \quad (S16)$$

Eq. (S16) leads to

$$a_{e_2}^{2\downarrow} = e^{2iq_{2,e,z}d_1}\left(\frac{D_1\varepsilon_{2,z}q_{2,e,z} - 1}{D_1\varepsilon_{2,z}q_{2,e,z} + 1}\right)a_{e_2}^{2\uparrow} \quad (S17)$$

Inserting Eq. (S17) into Eq. (S11), and dividing the two boundary conditions, we get to

$$\frac{a_{e_2}^{3\uparrow}e^{iq_{3,e,z}(d_1+d_2)} + a_{e_2}^{3\downarrow}e^{-iq_{3,e,z}(d_1+d_2)}}{\varepsilon_{2,z}q_{2,e,z}\left(a_{e_2}^{2\uparrow}e^{iq_{2,e,z}(d_1+d_2)} - a_{e_2}^{2\downarrow}e^{-iq_{2,e,z}(d_1+d_2)}\right)} = D_2 \quad (S18)$$

where

$$D_2 = \frac{D_1\varepsilon_{2,z}q_{2,e,z}cos(\xi_{2,e,z}) + isin(\xi_{2,e,z})}{\varepsilon_{2,z}q_{2,e,z}cos(\xi_{2,e,z}) + iD_1(\varepsilon_{2,z}q_{2,e,z})^2 sin(\xi_{2,e,z})} \quad (S19)$$

Eq. (S19) leads to

$$a_{e_2}^{3\downarrow} = e^{2iq_{3,e,z}(d_1+d_2)}\left(\frac{D_2\varepsilon_{3,z}q_{3,e,z} - 1}{D_2\varepsilon_{3,z}q_{3,e,z} + 1}\right)a_{e_2}^{3\uparrow} \quad (S20)$$

Inserting Eq. (S20) into Eq. (S10), and dividing the two boundary conditions, we get to

$$-\frac{1}{\varepsilon_4 q} = \frac{D_2\varepsilon_{3,z}q_{3,e,z}cos(\xi_{3,e,z}) + isin(\xi_{3,e,z})}{\varepsilon_{3,z}q_{3,e,z}cos(\xi_{3,e,z}) + iD_3(\varepsilon_{3,z}q_{3,e,z})^2 sin(\xi_{3,e,z})} \quad (S21)$$

Eq. (S21) can be rewritten as follows:

$$tan(\xi_{3,e,z}) = \frac{\dfrac{i}{D_2\varepsilon_{3,z}q_{3,e,z}} + \dfrac{i\varepsilon_4 q}{\varepsilon_{3,z}q_{3,e,z}}}{1 + \dfrac{\varepsilon_4 q}{D_2(\varepsilon_{3,z}q_{3,e,z})^2}} \quad (S22)$$

Using the identity $atan\left(\frac{x+y}{1-xy}\right) = atan(x) + atan(y)$ and noting that $\xi_{3,e,z} = \mp\frac{iqk_0d_3}{\rho_3}$ leads to the following equation describing the dispersion of highly confined polaritons in the investigated layer stack:

$$q = \frac{\rho_3}{k_0 d_3} \left[ atan\left(\frac{i\varepsilon_4 q}{\varepsilon_{3,z} q_{3,e,z}}\right) + atan\left(\frac{i}{D_2 \varepsilon_{3,z} q_{3,e,z}}\right) + \pi l \right] \text{ (S23)}$$

Noting that $q_{j,e,z} = \pm i \frac{q}{\rho_j}$, we can rewrite Eq. (S23) as follows:

$$q = \frac{\rho_3}{k_0 d_3} \left[ atan\left(\frac{\varepsilon_4 \rho_3}{\varepsilon_{3,z}}\right) + atan\left(\frac{\rho_3}{D_2 \varepsilon_{3,z} q}\right) + \pi l \right] \text{ (S24)}$$

with

$$D_2 = \frac{D_1 \varepsilon_{2,z} q_{2,e,z} cos(\xi_{2,e,z}) + isin(\xi_{2,e,z})}{\varepsilon_{2,z} q_{2,e,z} cos(\xi_{2,e,z}) + D_1 (\varepsilon_{2,z} q_{2,e,z})^2 isin(\xi_{2,e,z})}, \text{(S25)}$$

$$D_1 = \frac{D_0 \varepsilon_{1,z} q_{1,e,z} cos(\xi_{1,e,z}) + isin(\xi_{1,e,z})}{\varepsilon_{1,z} q_{1,e,z} cos(\xi_{1,e,z}) + D_0 (\varepsilon_{1,z} q_{1,e,z})^2 isin(\xi_{1,e,z})} \text{ (S26)}$$

and

$$D_0 = \frac{i}{\varepsilon_0 q} \text{ (S27)}$$

To obtain the out-of-plane Fresnel reflection coefficient, we add to Eq. (S10) a downgoing contribution $a_p^{4\downarrow} e^{iq_{4,p,z}(d_1+d_2+d_3)}$:

$$a_p^{4\uparrow} e^{iq_{4,p,z}(d_1+d_2+d_3)} + a_p^{4\downarrow} e^{iq_{4,p,z}(d_1+d_2+d_3)} = a_{e_2}^{3\uparrow} e^{iq_{3,e,z}(d_1+d_2+d_3)} + a_{e_2}^{3\downarrow} e^{-iq_{3,e,z}(d_1+d_2+d_3)}$$
$$-\varepsilon_4 q\left(a_p^{4\uparrow} e^{iq_{4,p,z}(d_1+d_2+d_3)} - a_p^{4\downarrow} e^{iq_{4,p,z}(d_1+d_2+d_3)}\right) = \varepsilon_{3,z} q_{3,e,z}\left(a_{e_2}^{3\uparrow} e^{iq_{3,e,z}(d_1+d_2+d_3)} - a_{e_2}^{3\downarrow} e^{-iq_{3,e,z}(d_1+d_2+d_3)}\right) \text{ (S28)}$$

Dividing the boundary conditions in Eq. (S28), we get

$$\frac{a_p^{4\uparrow} e^{iq_{4,p,z}(d_1+d_2+d_3)} + a_p^{4\downarrow} e^{-iq_{4,p,z}(d_1+d_2+d_3)}}{-\varepsilon_3 q\left(a_p^{4\uparrow} e^{iq_{4,p,z}(d_1+d_2+d_3)} - a_p^{4\downarrow} e^{iq_{4,p,z}(d_1+d_2+d_3)}\right)} = D_3 \quad \text{(S29)}$$

where

$$D_3 = \frac{D_2 \varepsilon_{3,z} q_{3,e,z} cos(\xi_{3,e,z}) + isin(\xi_{3,e,z})}{\varepsilon_{3,z} q_{3,e,z} cos(\xi_{3,e,z}) + D_2 (\varepsilon_{3,z} q_{3,e,z})^2 isin(\xi_{3,e,z})} \text{ (S30)}$$

The Fresnel reflection coefficient is found as the quotient

$$r_p = \frac{E_{4,r,z}}{E_{4,i,z}} \quad \text{(S31)}$$

where $E_{4,r,z} = a_p^{4\uparrow} e^{iq_{4,p,z}(d_1+d_2+d_3)}$ and $E_{4,i,z} = a_p^{4\downarrow} e^{-iq_{4,p,z}(d_1+d_2+d_3)}$. It reads, explicitly:

$$r_p = \frac{(D_3 \varepsilon_4 q - 1)}{(D_3 \varepsilon_4 q + 1)} \quad \text{(S32)}$$